\documentclass[onecolumn,a4paper,12pt,oneside]{article} 
\usepackage[top=2.5cm,left=2.5cm,right=2.5cm,bottom=3cm]{geometry}    
\usepackage[nostamp]{draftwatermark} % watermark programmed but not printed
\SetWatermarkText{\bf DRAFT}
\SetWatermarkAngle{45}
\SetWatermarkScale{4}
\SetWatermarkLightness{0.85}     
\usepackage[parfill]{parskip}    % Activate to begin paragraphs with an empty line rather than an indent
\usepackage{graphicx}
\usepackage{amssymb}
\usepackage{amsmath}
\usepackage{amsfonts}
\usepackage{amsthm}
\usepackage{epstopdf}
\usepackage[usenames,dvipsnames,svgnames,table]{xcolor}
\usepackage{tikz}
\usepackage{array}
\usepackage[T1]{fontenc}
\usepackage[utf8]{inputenc}
\usepackage{booktabs} % for much better looking tables
\usepackage{array} % for better arrays (eg matrices) in maths
\usepackage{paralist} % very flexible & customisable lists (eg. enumerate/itemize, etc.)
\usepackage{listings} 
\usepackage{verbatim} % adds environment for commenting out blocks of text & for better verbatim
\usepackage{subfig} % make it possible to include more than one captioned figure/table in a single float
\usepackage[colorlinks=true]{hyperref}
\hypersetup{colorlinks=true, linktocpage=true, linkcolor=Blue, citecolor=Green,
hypertexnames=true, urlcolor=Blue,pdftex}
\usepackage[english]{babel}
\usepackage{wrapfig}
\usepackage{multirow}
\usepackage{quoting}
\usepackage{rotating}
\quotingsetup{font=normalsize}
\theoremstyle{plain}

\usepackage{bm}
\usepackage{abstract}
\usepackage{cite}
\usepackage{float}
\usepackage{textcomp} 	
\usepackage{tabularx} % Better tables
\usepackage{caption}
\usepackage{authblk} % for authors, affiliations etc.
\usepackage{eso-pic} % for background figures
\usepackage{enumitem} % betters lists
\usepackage{lineno} % for numbering lines
\providecommand{\keywords}[1]{\textbf{{Key words: }} #1} % command for keywords

\usepackage{graphicx}
\usepackage{multirow}
\usepackage{amsmath,amssymb,amsfonts}
\usepackage{mathrsfs}
\usepackage[title]{appendix}
\usepackage{xcolor}
\usepackage{textcomp}
\usepackage{manyfoot}
\usepackage{booktabs}
\usepackage{algorithm}
\usepackage{algorithmicx}
\usepackage{algpseudocode}
\usepackage{listings}
\usepackage{paralist}

\newcommand{\be}{\begin{equation}}
\newcommand{\ee}{\end{equation}}
\newcommand{\bs}{\begin{split}}
\newcommand{\es}{\end{split}}
\newcommand{\bsig}{\boldsymbol{\sigma}}

\renewcommand{\Phi}{\varPhi}

\renewcommand{\Theta}{\varTheta}
\renewcommand{\Psi}{\varPsi}
\renewcommand{\Sigma}{\varSigma}

\renewcommand{\Delta}{\varDelta}
\renewcommand{\phi}{\varphi}
\renewcommand{\psi}{\varPsi}

\title{\LARGE{\bf Complete set of  bounds for the technical moduli in 3D anisotropic elasticity}}

\author{Paolo Vannucci\smallskip\\
\begin{small}{ LMV - Laboratoire de Mathématiques de Versailles, UMR8100 \\
 Université de Versailles et Saint Quentin - 45, Avenue des Etats-Unis, 78035 - France\\
           \href{mailto:paolo.vannucci@uvsq.fr}{paolo.vannucci@uvsq.fr}} \end{small} \bigskip\\
             - PREPRINT -  \bigskip\bigskip\\
             Final version in Journal of Elasticity \href{https://doi.org/10.1007/s10659-024-10062-z}{https://doi.org/10.1007/s10659-024-10062-z}}

\begin{document}
\maketitle

\begin{abstract}
The paper addresses the problem of finding the necessary and sufficient conditions to be satisfied by the engineering moduli of an anisotropic material for the elastic energy to be positive for each state of strain or stress. The problem is solved first in the most general case of a triclinic material and then each possible case of elastic syngony is treated as a special case. The method of analysis is based upon a rather forgotten theorem of linear algebra and, in the most general case, the calculations, too much involved, are carried out using a formal computation code. New, specific bounds, concerning some of the technical constants, are also found.

%\subclass{  74B05 \and 74E10  \and 74G45}
\end{abstract}

\section{Introduction}
\label{intro}
It is well known that the physical requirement for the work done on an elastic continuum to be positive, resumes to the circumstance that the elastic energy must be positive everywhere in the body for any possible deformation. This  fact determines a set of conditions for the moduli describing the material's response, the so-called {\it elastic bounds}. Such bounds define the admissible domain for a set of elastic parameters:  the Cartesian components of the elastic tensors (stiffness and compliance) or the technical constants (also known as {\it engineering moduli}) or, in the isotropic case, the Lamé's parameters. 
The determination of such bounds is hence a basic problem in the theory of classical elasticity, but rather surprisingly, left apart the case of isotropic materials, it has not yet received a complete solution.

In the literature, the determination of the elastic bounds is normally done by mechanical approaches, i.e. imposing the positiveness of the elastic energy for particularly simple states of strain or stress (e.g., a unidirectional or a shear state, or also a spherical  and a deviatoric state), and usually for isotropic materials. This is done, explicitly or implicitly,  in classical texts like, for instance and in the chronological order, the treatises of Love, \cite{Love}, Sokolnikoff, \cite{sokolnikoff}, Muskhelishvili, \cite{musk}, Mandel, \cite{mandel}, Baldacci, \cite{baldacci}, Gurtin, \cite{gurtin}.

Unlike the case of isotropic materials, the most general case of anisotropic solids has not received the same attention. The previous mentioned texts, e.g., limit their analysis to the isotropic case. The classical treatise of Lekhnitskii on anisotropic bodies, \cite{Lekhnitskii},  only partially considers the problem and, though introducing the technical constants named by himself {\it Chentsov's coefficients} and {\it coefficients of influence of the first or of the second type}, completely ignore the possible existence of some type of bounds for such parameters, limiting the analysis uniquely to  the Young's moduli, the shear moduli and the Poisson's ratios. More details are given in the recent book of Rand \& Rovenski, \cite{rand}, and especially in the technical text of Jones on composite materials, \cite{jones}. In all of these texts the authors make use of a mechanical approach,  like for isotropic materials, and only partial solutions are given, namely for the orthotropic case. Unlike in these texts, in the book of Nye on the physical properties of crystals, \cite{nye}, a mathematical approach is used, but it is not specified; in this way, omitting the procedure, the author gives some bounds for the Cartesian components of the stiffness elastic tensor for some crystal syngonies. A general mathematical approach, based on a rather forgotten theorem of linear algebra, is presented in the book of Ting, \cite{ting}.  However, the author does not develop the topic completely and he considers only the Cartesian components of the stiffness elastic tensor, not the technical constants.

The object of this paper is precisely to carry to its last consequences the approach suggested in the book of Ting: to find a complet set of bounds for the technical constants of an anisotropic material, i.e. the necessary and sufficient conditions  for the elastic potential to be positive, and this for the most general case of triclinic materials as well as for each one of the elastic syngonies. In the case of planar elasticity, this result has been recently obtained, \cite{vannucci23PRSA}, following an approach based upon the polar formalism, \cite{Verchery79,Meccanica05,vannucci_libro}. Previously, in \cite{vannucci15ijss}, the same problem has been solved in the set of the polar invariants of a planar material. This is not possible in three-dimensional elasticity, as the polar formalism applies only to plane problems. That is why in this paper we consider the technical constants, that, though frame-dependent quantities, are nonetheless the most used parameters  for the characterization of the elastic behavior of a material.

The paper is organized as follows: in the next Section, the basic equations are recalled. Then, the expression of the bounds in the general case of triclinic material are obtained and finally the cases corresponding to the different possible elastic syngonies are treated as special cases.

% anche nel caso isotropo, quante e quali sono le condizioni da scrivere, non è mai detto
  
\section{Basic relations}
Making use of  the Kelvin's  formalism, \cite{kelvin,kelvin1}, we represent the stress tensor $\bsig$ by a vector  $\{\sigma\}\in\mathbb{R}^6$
\be
\label{eq:sigma}
\{\sigma\}=\left\{\begin{array}{cccccc}\sigma_1,&\sigma_2,&\sigma_3,&\sigma_4,&\sigma_5,&\sigma_6\end{array}\right\}, 
\ee
 and the compliance tensor of elasticity $\mathbb{S}$ by a $6\times6$ symmetric matrix $[S]$:
 
 \be
 \label{eq:S}
 [S]=\left[
 \begin{array}{cccccc}
 S_{11}&S_{12}&S_{13}&S_{14}&S_{15}&S_{16}\\
 S_{12}&S_{22}&S_{23}&S_{24}&S_{25}&S_{26}\\
 S_{13}&S_{23}&S_{33}&S_{34}&S_{35}&S_{36}\\
 S_{14}&S_{24}&S_{34}&S_{44}&S_{45}&S_{46}\\
 S_{15}&S_{25}&S_{35}&S_{45}&S_{55}&S_{56}\\
 S_{16}&S_{26}&S_{36}&S_{46}&S_{56}&S_{66}
 \end{array}
 \right],
 \ee
with
\be
 \begin{array}{c}
 \sigma_i=\sigma_{ii},\ \sigma_p=\sqrt{2}\sigma_{hk},\medskip\\
 S_{ij}=\mathbb{S}_{iijj}, \ S_{ip}=\sqrt{2}\mathbb{S}_{iihk},\ S_{pq}=2\mathbb{S}_{ijhk},\medskip\\ \forall i,j,h,k\in\{1,2,3\},\ p,q\in\{4,5,6\};
 \end{array}
\ee
 the correspondences between the indices $p,q$ and the couples of indices $ij,$ $hk$ are
\be
1\rightarrow11,\ 2\rightarrow22,\ 3\rightarrow33,\ 4\rightarrow23,\ 5\rightarrow31,\ 6\rightarrow12.
\ee

We write the Cartesian components $S_{ij}$ as functions of the engineering constants, in order to find the bounds for these last. In the most general case of a completely anisotropic body, the 21 technical parameters are, cf.  \cite{Lekhnitskii,vannucci_libro}:
\begin{itemize}
\item  three Young's moduli $E_1,E_2,E_3$;\smallskip
\item  three shear moduli $G_{12},G_{23},G_{31}$;\smallskip
\item  three Poisson's coefficients $\nu_{12},\nu_{23},\nu_{31}$;\smallskip
\item  three Chentsov's coefficients $\mu_{12,23},\mu_{23,31},\mu_{31,12}$;\smallskip
\item nine coefficients of mutual influence of the first, $\eta_{i,jh}$,  or of the second type, $\eta_{ij,h},\ i,j,h=1,2,3$.
\end{itemize}

In the Kelvin's notation, the relations between the $S_{ij}$s and the technical constants are
\be
\label{eq:constants}
\begin{array}{c}
S_{11}=\dfrac{1}{E_1},\ S_{22}=\dfrac{1}{E_2},\ S_{33}=\dfrac{1}{E_3},\ S_{44}=\dfrac{1}{2G_{23}},\ S_{55}=\dfrac{1}{2G_{31}},\ S_{66}=\dfrac{1}{2G_{12}},\medskip\\
S_{12}=-\dfrac{\nu_{12}}{E_1}=-\dfrac{\nu_{21}}{E_2},\ S_{23}=-\dfrac{\nu_{23}}{E_2}=-\dfrac{\nu_{32}}{E_3},\ S_{31}=-\dfrac{\nu_{31}}{E_3}=-\dfrac{\nu_{13}}{E_1},\medskip\\
S_{14}=\dfrac{\eta_{1,23}}{\sqrt{2}G_{23}}=\dfrac{\eta_{23,1}}{\sqrt{2}E_1},
S_{15}=\dfrac{\eta_{1,31}}{\sqrt{2}G_{31}}=\dfrac{\eta_{31,1}}{\sqrt{2}E_1},
S_{16}=\dfrac{\eta_{1,12}}{\sqrt{2}G_{12}}=\dfrac{\eta_{12,1}}{\sqrt{2}E_1},\medskip\\
S_{24}=\dfrac{\eta_{2,23}}{\sqrt{2}G_{23}}=\dfrac{\eta_{23,2}}{\sqrt{2}E_2},
S_{25}=\dfrac{\eta_{2,31}}{\sqrt{2}G_{31}}=\dfrac{\eta_{31,2}}{\sqrt{2}E_2},
S_{26}=\dfrac{\eta_{2,12}}{\sqrt{2}G_{12}}=\dfrac{\eta_{12,2}}{\sqrt{2}E_2},\medskip\\
S_{34}=\dfrac{\eta_{3,23}}{\sqrt{2}G_{23}}=\dfrac{\eta_{23,3}}{\sqrt{2}E_3},
S_{35}=\dfrac{\eta_{3,31}}{\sqrt{2}G_{31}}=\dfrac{\eta_{31,3}}{\sqrt{2}E_3},
S_{36}=\dfrac{\eta_{3,12}}{\sqrt{2}G_{12}}=\dfrac{\eta_{12,3}}{\sqrt{2}E_3},\medskip\\
S_{45}=\dfrac{\mu_{23,31}}{2G_{31}}=\dfrac{\mu_{31,23}}{2G_{23}},
S_{46}=\dfrac{\mu_{12,23}}{2G_{23}}=\dfrac{\mu_{23,12}}{2G_{12}},
S_{56}=\dfrac{\mu_{31,12}}{2G_{12}}=\dfrac{\mu_{12,31}}{2G_{31}},
\end{array}
\ee
In  the above equations we have used the following {\it reciprocity relations}, \cite{jones,vannucci_libro}:
\be
\label{eq:poissons1}
\frac{\nu_{ij}}{E_i}=\frac{\nu_{ji}}{E_j},\ \ \ \frac{\mu_{ij,ki}}{G_{ki}}=\frac{\mu_{ki,ij}}{G_{ij}},\ \ \ \frac{\eta_{ij,k}}{E_k}=\frac{\eta_{k,ij}}{G_{ij}},\ \forall i,j=1,2,3,
\ee 

The elastic energy density
\be
V=\frac{1}{2}\{\sigma\}[S]\{\sigma\}^\top
\ee
is positive for each possible stress state $\{\sigma\}$ if and only if the matrix $[S]$ is positive definite. 
It is exactly this condition to give the bounds on the elastic moduli. Usually, this is done using a mechanical approach: because $V$ must be positive {\it for each possible stress state}, some particularly simple stress states are used. This  classical approach gives the entire set of elastic bounds for isotropic materials, but in the case of anisotropic elasticity two are the questions that remain unresolved in the literature: how many independent conditions exist (and hence: how many independent  stress states are to be considered) and how to write them in the most general case of complete anisotropy (triclinic materials). 
Actually, we know a set of conditions only for orthotropic materials and these conditions are written in the orthotropy frame, cf. \cite{jones}, but it is unclear wether or not such conditions constitute a complete set of independent bounds, i.e. whether they are necessary and sufficient conditions for $V$ to be positive for each possible stress state or not.

Another possible strategy is to use a mathematical approach: to write the  necessary and sufficient conditions for a symmetric matrix to be positive definite. In the most part of texts on continuum mechanics, and not only, this question is linked to a general result on eigenvalues ($[S]$ being symmetric, the spectral theorem, \cite{hohn58},  guarantees the existence of six real eigenvalues): a symmetric matrix is positive definite if and only if all of its eigenvalues are positive. Such theorem finds its deep interest in mechanics, where eigenvalues have specific physical meanings (principal stresses or strains, principal inertia moments, vibration frequencies, critical loads etc.). However, when general conditions are looked for, the analytical expression of the eigenvalues is needed. Unfortunately, the characteristic equation for $[S]$ is of degree six and, apart some specially simple cases, its solution is not known. 

There is, however, a much less known theorem of linear algebra giving a set of necessary and sufficient conditions for a real symmetric matrix to be positive definite. Quoting from Hohn, \cite{hohn58}:
{\it a real quadratic form is positive definite if and only if the leading principal minors of the matrix of the form are all positive.}

The leading principal minors of a $n\times n$ matrix are the  determinants of the $n$ sub-matrices obtained taking the $n$ first rows and columns, in the order. For $[S]$, the leading principal minors are:
\be
\label{eq:minors}
\begin{split}
&M_1{=}S_{11},\ M_2=\left|
 \begin{array}{cccccc}
 S_{11}&S_{12}\\
 S_{12}&S_{22} \end{array}
 \right|,\ M_3=\left|
 \begin{array}{cccccc}
 S_{11}&S_{12}&S_{13}\\
 S_{12}&S_{22}&S_{23}\\
 S_{12}&S_{23}&S_{33}
  \end{array}
 \right|,\ 
 M_4=\left|
 \begin{array}{cccccc}
 S_{11}&S_{12}&S_{13}&S_{14}\\
 S_{12}&S_{22}&S_{23}&S_{24}\\
 S_{12}&S_{23}&S_{33}&S_{34}\\
 S_{14}&S_{24}&S_{34}&S_{44}
 \end{array}
 \right|,\\
 &\ M_5=\left|
 \begin{array}{cccccc}
 S_{11}&S_{12}&S_{13}&S_{14}&S_{15}\\
 S_{12}&S_{22}&S_{23}&S_{24}&S_{25}\\
 S_{12}&S_{23}&S_{33}&S_{34}&S_{35}\\
 S_{14}&S_{24}&S_{34}&S_{44}&S_{45}\\
 S_{15}&S_{25}&S_{35}&S_{45}&S_{55}
 \end{array}
 \right|,\ 
 M_6=\det[S]=\left|
 \begin{array}{cccccc}
 S_{11}&S_{12}&S_{13}&S_{14}&S_{15}&S_{16}\\
 S_{12}&S_{22}&S_{23}&S_{24}&S_{25}&S_{26}\\
 S_{12}&S_{23}&S_{33}&S_{34}&S_{35}&S_{36}\\
 S_{14}&S_{24}&S_{34}&S_{44}&S_{45}&S_{46}\\
 S_{15}&S_{25}&S_{35}&S_{45}&S_{55}&S_{56}\\
 S_{16}&S_{26}&S_{36}&S_{46}&S_{56}&S_{66}
 \end{array}
 \right|.
\end{split}
\ee
By the above theorem, the necessary and sufficient conditions for $V$ to be positive for any stress state are 
\be
\label{eq:condMi}
M_i>0\ \forall i=1,...,6.
\ee
The true difference with respect to the method based upon the eigenvalues is that it is  possible to obtain an analytical expression for all  the minors $M_i$, though in the most general case of a triclinic material, i.e. with all the $S_{ij}s\neq0$ this is very complicate. The use of modern formal codes for mathematical computations can   reveal to be decisive in such a case, which has been done also for this research.

Two general considerations can be done. First, the above theorem ensures that the positiveness of $V$ can be stated with at most six necessary and sufficient conditions. However, we will see that the existence of  material symmetries generally reduces the number of these conditions. Second, the set of necessary and sufficient conditions for $V$ to be positive is not unique. In fact, the order of the Kelvin's vector $\{\sigma\}$ representing $\bsig$, eq. (\ref{eq:sigma}), is completely arbitrary: choosing another order modifies the rows of the matrix $[S]$, eq. (\ref{eq:S}), and by consequence all the minors $M_i$ in eq. (\ref{eq:minors}). There are 6!=720 different ways to order without repetitions the six components $\sigma_i$ into a set of 6 places. By consequence, there are 720 different, but equivalent, ways to write the necessary and sufficient conditions for the positiveness of $V$  for a triclinic material. A condition is however shared by  all of these 720 sets: the one on $M_6$, i.e. $\det[S]>0$.

We will use the usual Kelvin order for $\{\sigma\}$, i.e. the one in eq. (\ref{eq:sigma}), for writing the bounds on the technical constants. Then, in doing that, we will use:
\begin{itemize}
\item for $S_{ij},\ i,j=1,2,3,\ i\neq j$ the first one of their expression in eq. (\ref{eq:constants}), e.g. $S_{12}=-\dfrac{\nu_{12}}{E_1}$;\smallskip
\item for $S_{ip},\ i=1,2,3,\ p=4,5,6,$ their expression using the second coefficient of mutual influence, e.g. $S_{14}=\dfrac{\eta_{23,1}}{\sqrt{2}E_1}$ etc.;\smallskip
\item for $S_{pq},\ p,q=4,5,6,\ p\neq q$, the first one of their expression in eq. (\ref{eq:constants}), e.g. $S_{45}=\dfrac{\mu_{23,31}}{2G_{31}}$.
\end{itemize}
Successively, the expressions of the $M_i$s so obtained will be simplified using the reciprocity relations (\ref{eq:poissons1}), so as to express each $M_i$ as the product of a function depending exclusively on the moduli $E_i$s and $G_{ij}$s, $i,j=1,2,3$, and of another a function of only the coefficients, $\nu_{ij}$s, $\eta_{i,jk}$s, $\eta_{jk,i}$s, $\mu_{ij,ki}$s and $\mu_{ki,ij}$s, $i,j,k=1,2,3$.

\section{The general case: bounds for a triclinic material}
The  case of triclinic materials is presented. This is the most complicate case, giving rise to very long expressions for $M_5$ and $M_6$, given here for the sake of completeness, and obtained by a formal computation code and then simplified by hand. 

\subsection{Necessary and sufficient conditions}
The $M_i$s, eq. (\ref{eq:minors}), once used eqs. (\ref{eq:poissons1}) and after simplification, are: 
\be
\label{eq:triclinic1}
M_1>0\rightarrow \dfrac{1}{E_1}>0,%\smallskip
\ee
\be
\label{eq:triclinic2}
M_2>0\rightarrow\dfrac{1-\nu_{12}\ \nu_{21}}{E_1E_2}>0,% \smallskip
\ee
\be
\label{eq:triclinic3}
M_3>0\rightarrow
\dfrac{1- \nu_{12}\ \nu_{21} -\nu_{23}\ \nu_{32}- \nu_{31} \ \nu_{13}    -2 \nu_{12}\ \nu_{23}\ \nu_{31}}{E_1E_2E_3}>0,%\smallskip
\ee
\be
\label{eq:triclinic4}
 \begin{array}{ll}M_4>0\rightarrow&
\dfrac{1}{E_1E_2E_3G_{23}}\{1-\nu_{12}\ \nu_{21}-\nu_{23}\ \nu_{32}-\nu_{31}\ \nu_{13}-2\nu_{12}\ \nu_{23}\ \nu_{31}+\\
&+\nu_{12}\ \nu_{21}\ \eta_{3,23}\ \eta_{23,3}+\nu_{23}\ \nu_{32}\ \eta_{1,23}\ \eta_{23,1}+\nu_{31}\ \nu_{13}\ \eta_{2,23}\ \eta_{23,2}-\\
&-\eta_{1,23}\left[\eta_{23,1}+2\nu_{12}(\eta_{23,2}+\eta_{23,3}\nu_{23})\right]-\\
&-\eta_{2,23}\left[\eta_{23,2}+2\nu_{23}(\eta_{23,3}+\eta_{23,1}\nu_{31})\right]-\\
&-\eta_{3,23}\left[\eta_{23,3}+2\nu_{31}(\eta_{23,1}+\eta_{23,2}\nu_{12})\right]\}>0,\smallskip
\end{array}
\ee
\be
\label{eq:triclinic5}
\begin{array}{ll}M_5>0\rightarrow
&\dfrac{1}{E_1E_2E_3G_{23}G_{31}}
\{[\eta_{3,23} (\eta_{23,3}-2 \eta_{31,3} \mu_{23,31})+\\
&+ (\eta_{31,3} \eta_{3,31}\hspace{-.7mm}+\hspace{-.7mm}\mu_{23,31} \mu_{31,23}\hspace{-.7mm}-\hspace{-.7mm}1))]\nu_{12} \nu_{21}\hspace{-.7mm}+ \hspace{-.7mm}(\eta_{3,23} \eta_{1,31}\hspace{-.7mm}-\hspace{-.7mm}\eta_{1,23} \eta_{3,31}) \times\\
&\times[\eta_{23,3} (\eta_{31,1}+2 \eta_{31,2} \nu_{12})]-{\eta_{31,3} (\eta_{23,1}+2 \eta_{23,2} \nu_{12})- }\\
&-{[ \eta_{1,23} \eta_{23,1}+ \eta_{1,31} (\eta_{31,1}+2 \nu_{12} (\eta_{31,2}+\eta_{31,3} \nu_{23}))-}\\
&-2 (\eta_{1,23} \eta_{31,1} \mu_{23,31}+\eta_{1,31} \mu_{31,23} \nu_{12} (\eta_{23,2}+\eta_{23,3} \nu_{23})+\\
&+\eta_{1,23} \nu_{12} (\eta_{31,2} \mu_{23,31}-\eta_{23,3} \nu_{23}+\eta_{31,3} \mu_{23,31} \nu_{23}-\eta_{23,2}))]+\\
&+[ \eta_{2,23}(\eta_{23,2}\hspace{-.7mm}-\hspace{-.7mm}2 \eta_{31,2} \mu_{23,31})\hspace{-.7mm} +\hspace{-.7mm}\eta_{2,31} \eta_{31,2}\hspace{-.7mm}+\hspace{-.7mm}\mu_{23,31} \mu_{31,23}-1] \hspace{-.3mm}\nu_{13} \nu_{31}+ \\
&+(\eta_{2,23} \eta_{1,31}\hspace{-.7mm}-\hspace{-.7mm}\eta_{1,23} \eta_{2,31}) [\eta_{23,2} \eta_{31,1}+2 \eta_{23,3} \eta_{31,1} \nu_{23}\hspace{-.7mm}-\hspace{-.7mm}\eta_{23,1} (\eta_{31,2}+\\
&+2 \eta_{31,3} \nu_{23})]+ \eta_{1,31} \eta_{31,1} \nu_{23} \nu_{32}+ \eta_{1,23} (\eta_{23,1}-2 \eta_{31,1} \mu_{23,31})\times\\
&\times \nu_{23} \nu_{32}-(\eta_{2,23} \eta_{3,31}-\eta_{3,23} \eta_{3,31}) [\eta_{23,3} \eta_{31,2}+2 \eta_{23,1} \eta_{31,2} \nu_{31}-\\
&-\eta_{23,2} (\eta_{31,3}\hspace{-.7mm}+\hspace{-.7mm}2 \eta_{31,1} \nu_{31})]\hspace{-.7mm}-\hspace{-.7mm} [\eta_{2,31}(\eta_{31,2}\hspace{-.7mm}+\hspace{-.7mm}2 \nu_{23} (\eta_{31,3}+\eta_{31,1} \nu_{31})]-\\
&-2[\eta_{2,23}\eta_{31,2} \mu_{23,31}+\eta_{2,31} \mu_{31,23} \nu_{23} (\eta_{23,3}+\eta_{23,1} \nu_{31})+\\
&+\eta_{2,23} \nu_{23} (\eta_{31,3} \mu_{23,31}-\eta_{23,1} \nu_{31}+\eta_{31,1} \mu_{23,31} \nu_{31}-\eta_{23,3}))]-\\
&-[ \eta_{3,31} (\eta_{31,3}+2 (\eta_{31,1}+\eta_{31,2} \nu_{12}) \nu_{31})-2{  }\mu_{23,31}\eta_{3,31}(\eta_{23,3}+\\
&+ (\eta_{23,1}+\eta_{23,2} \nu_{12}) \nu_{31}+\eta_{23,3} (\eta_{31,1}\mu_{23,31}-\eta_{23,2} \nu_{12}+\\
&+\eta_{31,2} \mu_{23,31} \nu_{12}-\eta_{23,1}) \nu_{31})+(\eta_{23,2} \eta_{2,23}+\eta_{23,3} \eta_{3,23}-\\
&-(1-\mu_{23,31} \mu_{31,23}) (1-2 \nu_{12}\nu_{23} \nu_{31}-\nu_{23} \nu_{32}))]\}>0,
\end{array}
\ee

\be
\label{eq:triclinic6}
\begin{array}{ll}M_6>0\rightarrow&\dfrac{1}{E_1E_2E_3G_{12}G_{23}G_{31}}
\{1{-}\eta_{12,1} \eta_{1,12}{-}\eta_{1,23} \eta_{23,1}{-}\eta_{12,2} \eta_{2,12}{+}\\
&{+}\eta_{12,2}\eta_{1,23} \eta_{23,1} \eta_{2,12}{-}2 \eta_{12,2} \eta_{1,12} \eta_{23,1} \eta_{2,23}{-}\eta_{23,2} \eta_{2,23}{+}\\
&{+}\eta_{12,1} \eta_{1,12} \eta_{23,2} \eta_{2,23}{-}\eta_{1,31} \eta_{31,1}{+}{+}\eta_{12,2} \eta_{1,31} \eta_{2,12} \eta_{31,1}{+}\\
&{+}\eta_{1,31} \eta_{23,2} \eta_{2,23} \eta_{31,1}{-}2 \eta_{12,2} \eta_{1,12} \eta_{2,31} \eta_{31,1}{-}2 \eta_{1,23} \eta_{23,2} \eta_{2,31} \eta_{31,1}{-}\\
&{-}\eta_{2,31} \eta_{31,2}{+}\eta_{12,1} \eta_{1,12} \eta_{2,31} \eta_{31,2}{+}\eta_{1,23} \eta_{23,1} \eta_{2,31} \eta_{31,2}{-}\eta_{12,3} \eta_{3,12}{+}\\
&{+}\eta_{12,3} \eta_{23,2} \eta_{2,23} \eta_{3,12}{-}\eta_{12,3} \eta_{1,31} \eta_{23,2} \eta_{2,23} \eta_{31,1} \eta_{3,12}{+}2 \eta_{12,3} \eta_{1,31} \times\\
&\times\eta_{23,1} \eta_{2,23} \eta_{31,2} \eta_{3,12}{-}2 \eta_{12,1} \eta_{1,31} \eta_{23,3} \eta_{2,23} \eta_{31,2} \eta_{3,12}{+}\eta_{12,3} \eta_{2,31} \times\\
&\times\eta_{31,2} \eta_{3,12}{-}\eta_{12,3} \eta_{1,23} \eta_{23,1} \eta_{2,31} \eta_{31,2} \eta_{3,12}{-}\eta_{23,3} \eta_{3,23}{-}\\
&{-}2 \eta_{12,3} \eta_{23,2} \eta_{2,12} \eta_{3,23}{+}\eta_{12,2} \eta_{23,3} \eta_{2,12} \eta_{3,23}{+}2 \eta_{12,3} \eta_{1,31} \eta_{23,2} \times\\
&\times\eta_{2,12} \eta_{31,1} \eta_{3,23}{-}\eta_{12,2} \eta_{1,31} \eta_{23,3} \eta_{2,12} \eta_{31,1} \eta_{3,23}{+}2 \eta_{12,2} \eta_{1,12}\times\\
&\times \eta_{23,3} \eta_{2,31} \eta_{31,1} \eta_{3,23}{-}2 \eta_{12,2} \eta_{1,31} \eta_{23,1} \eta_{2,12} \eta_{31,2} \eta_{3,23}{+}\\
&{+}2 \eta_{12,3} \eta_{1,12} \eta_{23,1} \eta_{2,31} \eta_{31,2} \eta_{3,23}{+}\eta_{23,3} \eta_{2,31} \eta_{31,2} \eta_{3,23}{-}\\
&{-}\eta_{12,1} \eta_{1,12} \eta_{23,3} \eta_{2,31} \eta_{31,2} \eta_{3,23}{-}2 \eta_{12,2} \eta_{1,12} \eta_{23,1} \eta_{2,31} \eta_{31,3} \eta_{3,23}{-}\\
&{-}2 \eta_{12,3} \eta_{1,23} \eta_{23,2} \eta_{2,12} \eta_{31,1} \eta_{3,31}{+}2 \eta_{12,2} \eta_{1,23} \eta_{23,3} \eta_{2,12} \eta_{31,1} \eta_{3,31}{+}\\
&{+}2 \eta_{12,3} \eta_{1,12} \eta_{23,2} \eta_{2,23} \eta_{31,1} \eta_{3,31}{-}2 \eta_{12,2} \eta_{1,12} \eta_{23,3} \eta_{2,23} \eta_{31,1} \eta_{3,31}{-}\\
&{-}2 \eta_{12,3} \eta_{2,12} \eta_{31,2} \eta_{3,31}{+}2 \eta_{12,3} \eta_{1,23} \eta_{23,1} \eta_{2,12} \eta_{31,2} \eta_{3,31}{-}\\
&{-}2 \eta_{12,3} \eta_{1,12} \eta_{23,1} \eta_{2,23} \eta_{31,2} \eta_{3,31}{-}2 \eta_{23,3} \eta_{2,23} \eta_{31,2} \eta_{3,31}{+}\\
&{+}2 \eta_{12,1} \eta_{1,12} \eta_{23,3} \eta_{2,23} \eta_{31,2} \eta_{3,31}{-}\eta_{31,3} \eta_{3,31}{+}\eta_{12,2} \eta_{2,12} \eta_{31,3} \eta_{3,31}{-}\\
&{-}\eta_{12,2} \eta_{1,23} \eta_{23,1} \eta_{2,12} \eta_{31,3} \eta_{3,31}{+}2 \eta_{12,2} \eta_{1,12} \eta_{23,1} \eta_{2,23} \eta_{31,3} \eta_{3,31}{+}\\
&{+}\eta_{23,2} \eta_{2,23} \eta_{31,3} \eta_{3,31}{-}\eta_{12,1} \eta_{1,12} \eta_{23,2} \eta_{2,23} \eta_{31,3} \eta_{3,31}{+}\\
&{+}2 \eta_{23,2} \eta_{2,12} \mu_{12,23}{-}2 \eta_{1,31} \eta_{23,2} \eta_{2,12} \eta_{31,1} \mu_{12,23}{+}2 \eta_{1,12} \eta_{23,2} \eta_{2,31} \times\\
&\times\eta_{31,1} \mu_{12,23}{+}2 \eta_{1,31} \eta_{23,1} \eta_{2,12} \eta_{31,2} \mu_{12,23}{-}2 \eta_{1,12} \eta_{23,1} \eta_{2,31}\times\\ 
&\times\eta_{31,2} \mu_{12,23}{+}2 \eta_{23,3} \eta_{3,12} \mu_{12,23}{-}2 \eta_{23,3} \eta_{2,31} \eta_{31,2} \eta_{3,12}\mu_{12,23}{-}2 \eta_{23,2} \eta_{2,12}\times\\
&\times \eta_{3,31} \mu_{12,23}{+}2 \eta_{23,3} \eta_{2,12} \eta_{31,2} \eta_{3,31} \mu_{12,23}{+}2 \eta_{23,2} \eta_{31,2} \eta_{3,12} \eta_{3,31} \mu_{12,23}{+}\\
&+(\eta_{1,31} \eta_{3,12}{-}\eta_{1,12} \eta_{3,31}) (\eta_{12,3} \eta_{31,1}{-}\eta_{12,1} \eta_{31,3}{-}2 \eta_{23,3} \eta_{31,1} \mu_{12,23}{+}\\
&{+}2 \eta_{23,1} \eta_{31,3} \mu_{12,23}){{-}}\mu_{12,23} \mu_{23,12}{+}\eta_{2,31} \eta_{31,2} \mu_{12,23} \mu_{23,12}{+}\eta_{31,3} \eta_{3,31} \mu_{12,23} \mu_{23,12}{-}\\
&{-}2 \eta_{12,2} \eta_{1,23} \eta_{2,12} \eta_{31,1} \mu_{23,31}{+}2 \eta_{12,2} \eta_{1,12} \eta_{2,23} \eta_{31,1} \mu_{23,31}{+}2 \eta_{12,1} \eta_{1,23} \eta_{2,12} \eta_{31,2}\times\\
&\times \mu_{23,31}{+}2 \eta_{2,23} \eta_{31,2} \mu_{23,31}{-}2 \eta_{12,1} \eta_{1,12} \eta_{2,23} \eta_{31,2} \mu_{23,31}{-}2 \eta_{12,3} \eta_{
2,23} \eta_{31,2} \eta_{3,12} \mu_{23,31}{+}\\
&+2 \eta_{12,2} \eta_{2,23} \eta_{31,3} \eta_{3,12} \mu_{23,31}{+}2 \eta_{12,3} \eta_{2,12} \eta_{31,2} \eta_{3,23} \mu_{23,31}{+}2 \eta_{31,3} \eta_{3,23}
\mu_{23,31}{-}\\
&-2 \eta_{12,2} \eta_{2,12} \eta_{31,3} \eta_{3,23} \mu_{23,31}{-}2 \eta_{2,12} \eta_{31,2} \mu_{12,23} \mu_{23,31}{-}2 \eta_{31,3} \eta_{3,12} \mu_{12,23} \mu_{23,31}{+}\\
&+(\eta_{1,23} \eta_{3,12}{-}\eta_{1,12} \eta_{3,23}) (\eta_{12,3} \eta_{23,1}{-}\eta_{12,1} \eta_{23,3}{-}2 \eta_{12,3} \eta_{31,1} \mu_{23,31}{+}2 \eta_{12,1}\times\\
&\times \eta_{31,3} \mu_{23,31}){+}2 \eta_{12,2} \eta_{1,31} \eta_{23,1}\eta_{2,23} \mu_{31,12}{-}2 \eta_{12,1} \eta_{1,31} \eta_{23,2} \eta_{2,23} \mu_{31,12}{+}\\
&+2 \eta_{12,2} \eta_{2,31} \mu_{31,12}{-}2 \eta_{12,2} \eta_{1,23} \eta_{23,1} \eta_{2,31} \mu_{31,12}{+}2 \eta_{12,1} \eta_{1,23} \eta_{23,2} \eta_{2,31} \mu_{31,12}{+}\\
&+2 \eta_{12,3} \eta_{23,2} \eta_{2,31} \eta_{3,23} \mu_{31,12}{-}2 \eta_{12,2} \eta_{23,3} \eta_{2,31} \eta_{3,23} \mu_{31,12}{+}2 \eta_{12,3} \eta_{3,31} \mu_{31,12}{-}\\
&-2 \eta_{12,3} \eta_{23,2} \eta_{2,23} \eta_{3,31} \mu_{31,12}{+}2 \eta_{12,2} \eta_{23,3} \eta_{2,23} \eta_{3,31} \mu_{31,12}{-}2 \eta_{23,2} \eta_{2,31} \mu_{12,23} \mu_{31,12}{-}\\
&-2 \eta_{23,3} \eta_{3,31} \mu_{12,23} \mu_{31,12}{-}\mu_{12,31} \mu_{31,12}{+}\eta_{23,2} \eta_{2,23} \mu_{12,31} \mu_{31,12}{+}\eta_{23,3} \eta_{3,23} \times\\
&\times\mu_{12,31} \mu_{31,12}{-}2 \eta_{12,2} \eta_{2,23} \mu_{23,31} \mu_{31,12}{-}2 \eta_{12,3} \eta_{3,23} \times\\
&\times\mu_{23,31}\mu_{31,12}{+}(\eta_{1,31} \eta_{3,23}{-}\eta_{1,23} \eta_{3,31}) (\eta_{23,3} \eta_{31,1}{-}\eta_{23,1} \eta_{31,3}{+}2 \eta_{12,3} \eta_{23,1} \mu_{31,12}{-}\\
&-2 \eta_{12,1} \eta_{23,3} \mu_{31,12}){-}\mu_{23,31} \mu_{31,23}{+}\eta_{12,2} \eta_{2,12} \mu_{23,31} \mu_{31,23}{+}\eta_{12,3} \eta_{3,12} \mu_{
23,31} \mu_{31,23}{+}\\
&continued...
\end{array}
\end{equation}
\begin{equation*}
\begin{array}{l}
...continued\\
+2 [({-}2 \eta_{1,31} \eta_{23,3} \eta_{31,2} \eta_{3,12}{+}\eta_{1,31} \eta_{23,2}\eta_{31,3} \eta_{3,12}{+}\eta_{1,12} \eta_{23,3} \eta_{31,2} \eta_{3,31}{-}\\
-\eta_{1,12} \eta_{23,2} \eta_{31,3} \eta_{3,31}) \mu_{12,23}{+}\eta_{12,2} \eta_{31,3} ({-}\eta_{1,31} \eta_{3,12}{+}\eta_{1,12} \eta_{3,31}{+}\eta_{1,31} \eta_{3,23} \mu_{23,12}{-}\\
-\eta_{1,23} \eta_{3,31} \mu_{23,12}){+}\eta_{12,3} \eta_{31,2} (\eta_{1,31} \eta_{3,12}{-}\eta_{1,12} \eta_{3,31}{+}\eta_{1,23} \eta_{3,31} \mu_{
23,12})] \nu_{12}{+}\\
+2 \{({-}\eta_{1,12} \eta_{23,3} \eta_{31,2}{+}\eta_{1,12} \eta_{23,2} \eta_{31,3}{+}\eta_{
23,1} \eta_{31,2} \eta_{3,12}) \eta_{3,23} \mu_{12,31}{+}\\
+\eta_{1,23} \eta_{3,31} ({-}\eta_{23,3} \eta_{31,2}{+}\eta_{23,2} \eta_{31,3}{-}2 \eta_{12,3} \eta_{23,2} \mu_{31,12}{+}\eta_{12,2} \eta_{23,3} \mu_{
31,12}){+}\\
+\eta_{1,31} \eta_{3,23} [\eta_{12,3} \eta_{23,2} \mu_{31,12}{-}\eta_{23,2} \eta_{31,3}{+}\eta_{23,3} (\eta_{31,2}{-}\eta_{12,2} \mu_{31,12})]\} \nu_{12}{+}\\
+2 [\eta_{12,3} (\eta_{1,23} \eta_{3,12}{-}\eta_{1,12} \eta_{3,23}) (\eta_{23,2}{-}\eta_{31,2} \mu_{23,31}){+}\eta_{12,2} \eta_{1,23} \eta_{3,12} (\eta_{31,3} \mu_{23,31}-\\
{-}\eta_{23,3}){+}\eta_{12,2} \eta_{1,31} \eta_{23,3} \eta_{3,12} \mu_{31,23}{+}\eta_{12,3} \eta_{23,2} ({-}\eta_{1,31} \eta_{3,12}{+}\eta_{1,12} \eta_{3,31}) \mu_{31,23}{+}\\
+\eta_{12,2} \eta_{1,12} \eta_{23,3} (\eta_{3,23}{-}2 \eta_{3,31} \mu_{31,23})] \nu_{12}{+}[{-}(1{-}\eta_{31,3} \eta_{3,31})(1{-}\mu_{12,23} \mu_{23,12}){+}\\
+\mu_{12,31} \mu_{31,12}{+}\eta_{23,3} (\eta_{3,23}{+}2 \eta_{3,31} \mu_{12,23} \mu_{31,12}{-}\eta_{3,23} \mu_{12,31} \mu_{31,12}){+}\\
+\mu_{23,31} (\mu_{31,23}{-}2 \eta_{31,3} \eta_{3,23}{+}2\eta_{31,3} \eta_{3,12} \mu_{12,23}{-}2 \mu_{12,23} \mu_{31,12}){+}\eta_{12,3} (\eta_{3,12}-\\
{-}2 \eta_{3,23} \mu_{23,12}{-}2 \eta_{3,31} \mu_{31,12}{+}2 \eta_{3,23} \mu_{23,31} \mu_{31,12}{-}\eta_{3,12} \mu_{23,31} \mu_{31,23})] \nu_{12} \nu_{21}-\\
{-}2 \eta_{12,3} \eta_{2,12} \nu_{23}{-}2 \eta_{23,3} \eta_{2,23} \nu_{23}{-}2 \eta_{2,31} \eta_{31,3} \nu_{23}{+}2 \eta_{23,3} \eta_{2,12} \mu_{12,23} \nu_{23}{+}\\
+2 \eta_{2,12} \eta_{31,3} \mu_{12,31} \nu_{23}{+}2 \eta_{12,3} \eta_{2,23} \mu_{23,12} \nu_{23}{+}2 \eta_{2,31} \eta_{31,3} \mu_{12,23} \mu_{23,12} \nu_{23}{-}\\
-2 \eta_{2,23} \eta_{31,3} \mu_{12,31} \mu_{23,12} \nu_{23}{+}2 \eta_{2,23} \eta_{31,3} \mu_{23,31} \nu_{23}{-}2 \eta_{2,12} \eta_{31,3} \mu_{12,23} \mu_{23,31} \nu_{23}{+}\\
+2 \eta_{12,3} \eta_{2,31} \mu_{31,12} \nu_{23}{-}2 \eta_{23,3} \eta_{2,31} \mu_{12,23} \mu_{31,12} \nu_{23}{+}2 \eta_{23,3} \eta_{2,23} \mu_{12,31} \mu_{31,12} \nu_{23}{-}\\
-2 \eta_{12,3} \eta_{2,23} \mu_{23,31} \mu_{31,12} \nu_{23}{+}2 \eta_{23,3} \eta_{2,31} \mu_{31,23} \nu_{23}{-}2 \eta_{23,3} \eta_{2,12} \mu_{12,31} \mu_{31,23} \nu_{23}{-}\\
-2 \eta_{12,3} \eta_{2,31} \mu_{23,12} \mu_{31,23} \nu_{23}{+}2 \eta_{12,3} \eta_{2,12} \mu_{23,31} \mu_{31,23} \nu_{23}{+}2 [(\eta_{12,3} \eta_{23,1}{-}\eta_{12,1} \eta_{23,3})\times\\
\times (\eta_{1,23} \eta_{2,12}{-}\eta_{1,12} \eta_{2,23}){+}(\eta_{1,31} \eta_{2,12}{-}\eta_{1,12} \eta_{2,31}) (\eta_{12,3} \eta_{31,1}{-}\eta_{12,1} \eta_{31,3}){+}\\
+(\eta_{1,31} \eta_{2,23}{-}\eta_{1,23} \eta_{2,31}) (\eta_{23,3} \eta_{31,1}{-}\eta_{23,1} \eta_{31,3}){+}\eta_{12,1} \eta_{1,31} \eta_{23,3} \eta_{2,12} \mu_{12,23}{+}\\
+\eta_{23,3} (\eta_{1,12} \eta_{2,31}{-}\eta_{1,31} \eta_{2,12}) \eta_{31,1} \mu_{12,23}{+}\eta_{23,1} (\eta_{1,31} \eta_{2,12}{-}2 \eta_{12,1} \eta_{31,2}) \eta_{31,3} \mu_{12,23}{+}\\
+\eta_{1,12} \eta_{23,1} \eta_{2,23} \eta_{31,3} \mu_{12,31}{+}\eta_{1,23} \eta_{2,12} (\eta_{23,3} \eta_{31,1}{-}\eta_{23,1} \eta_{
31,3}) \mu_{12,31}{+}\\
+\eta_{12,3} ({-}\eta_{1,31} \eta_{2,23}{+}\eta_{1,23} \eta_{2,31}) \eta_{31,1} \mu_{23,12}{+}\eta_{12,1} \eta_{1,31} \eta_{2,23} \eta_{31,3} \mu_{23,12}{+}\eta_{1,12} \eta_{2,23}\times\\
\times (\eta_{12,3} \eta_{31,1}{-}\eta_{12,1} \eta_{31,3}) \mu_{23,31}{+}\eta_{1,23} \eta_{2,12} (\eta_{12,1} \eta_{31,3}{-}2 \eta_{12,3}\eta_{31,1}) \mu_{23,31}{+}\\
+\eta_{12,3} \eta_{23,1} (\eta_{1,31} \eta_{2,23}{-}\eta_{1,23} \eta_{2,31}) \mu_{31,12}{+}\eta_{12,1} \eta_{23,3} (\eta_{1,23} \eta_{2,31}{-}2 \eta_{1,31} \eta_{2,23}) \mu_{31,12}{+}\\
+\eta_{1,12} (\eta_{12,3} \eta_{23,1}{-}\eta_{12,1} \eta_{23,3}) \eta_{2,31} \mu_{31,23}] \nu_{23}{+}2 \mu_{23,12}\nu_{12} (\eta_{1,23}{-}\eta_{1,31} \mu_{23,12}) \times\\
\times (\eta_{12,2}{+}\eta_{12,3} \nu_{23}){+}2 (\eta_{1,31}{-}\eta_{1,12} \mu_{12,31}) \mu_{31,23} \nu_{12} (\eta_{23,2}{+}\eta_{23,3} \nu_{23}){+}2 \mu_{12,31} \nu_{12}\times\\
\times(\eta_{1,12}{-}\eta_{1,23} \mu_{23,12})  (\eta_{31,2}{+}\eta_{31,3} \nu_{23}){+}\eta_{1,12} \mu_{23,31} \mu_{31,23} [\eta_{12,1}{+}2 \nu_{12} (\eta_{12,2}{+}\\
+\eta_{12,3} \nu_{23})]{+}\eta_{1,23} \mu_{12,31} \mu_{31,12} [\eta_{23,1}{+}2 \nu_{12} (\eta_{
23,2}{+}\eta_{23,3} \nu_{23})]{+}\eta_{1,31} \mu_{12,23} \mu_{23,12} \times\\
\times[\eta_{31,1}{+}2 \nu_{12} (\eta_{31,2}{+}\eta_{31,3} \nu_{23})]{+}2 \eta_{1,12} \{\eta_{23,1} \mu_{12,23}{-}\eta_{31,1} \mu_{12,23} \mu_{23,31}{-}\nu_{12}\times\\
\times [\eta_{12,2}{-}\eta_{23,2} \mu_{12,23}{+}\eta_{31,2} \mu_{12,23} \mu_{23,31}{+}(\eta_{
12,3}{-}\eta_{23,3} \mu_{12,23}{+}\eta_{31,3} \mu_{12,23} \mu_{23,31}) \nu_{23}]\}  {+}\\
+2 \eta_{1,31} \{\eta_{12,1} \mu_{31,12}{-}\eta_{23,1} \mu_{12,23} \mu_{31,12}{-}\nu_{12} [\eta_{31,2}{-}\eta_{12,2} \mu_{
31,12}{+}\eta_{23,2} \mu_{12,23} \mu_{31,12}{+}\\
+(\eta_{31,3}{-}\eta_{12,3} \mu_{31,12}{+}\eta_{23,3} \mu_{12,23} \mu_{31,12}) \nu_{23}]\}{+}2 \eta_{1,23} (\eta_{31,1} \mu_{23,31}{-}\eta_{12,1} \mu_{23,31} \mu_{31,12}{-}\\ 
-\nu_{12} \{\eta_{23,2}{-}\eta_{31,2} \mu_{23,31}{+}\eta_{12,2} \mu_{23,31} \mu_{31,12}{+}[\eta_{
23,3}{-}\eta_{31,3} \mu_{23,31}{+}\eta_{12,3} \mu_{23,31} \mu_{31,12})\times\\
\times \nu_{23}]\}{-}2 \eta_{12,1} \eta_{3,12} \nu_{31}{-}2 \eta_{12,2} \eta_{23,1} \eta_{2,23} \eta_{3,12} \nu_{31}{+}2 \eta_{12,1} \eta_{23,2} \eta_{2,23} \eta_{3,12} \nu_{31}-\\
{-}2 \eta_{12,2} \eta_{2,31} \eta_{31,1} \eta_{3,12} \nu_{31}{+}2 \eta_{12,1} \eta_{2,31} \eta_{31,2} \eta_{3,12} \nu_{31}{-}2 \eta_{23,1} \eta_{3,23} \nu_{31}{+}\\
+2 \eta_{12,2} \eta_{23,1} \eta_{2,12} \eta_{3,23} \nu_{31}{-}2 \eta_{12,1} \eta_{23,2} \eta_{2,12} \eta_{3,23} \nu_{31}{-}2 \eta_{23,2} \eta_{2,31} \eta_{31,1} \eta_{3,23} \nu_{31}+\\
{+}2 \eta_{23,1} \eta_{2,31} \eta_{31,2} \eta_{3,23} \nu_{31}{-}2 \eta_{31,1} \eta_{3,31} \nu_{31}{+}2 \eta_{12,2} \eta_{2,12} \eta_{31,1} \eta_{3,31} \nu_{31}{+}2 \eta_{23,2} \eta_{2,23}\times\\
\times \eta_{31,1} \eta_{3,31} \nu_{31}{-}2 \eta_{12,1} \eta_{2,12} \eta_{31,2} \eta_{3,31} \nu_{31}{-}2 \eta_{23,1} \eta_{2,23} \eta_{31,2} \eta_{3,31} \nu_{31}{+}2 \eta_{23,1} \eta_{3,12} \times\\
\times\mu_{12,23} \nu_{31}{+}2 \eta_{23,2} \eta_{2,31} \eta_{31,1} \eta_{3,12} \mu_{12,23} \nu_{31}{-}2 \eta_{23,1} \eta_{2,31} \eta_{31,2} \eta_{3,12} \mu_{12,23} \nu_{31}{+}\\
+2 \eta_{23,1} \eta_{2,12} \eta_{31,2} \eta_{3,31} \mu_{12,23} \nu_{31}{-}4 \eta_{23,2} \eta_{31,1} \eta_{3,12} \eta_{3,31} \mu_{12,23} \nu_{31}{+}2 \eta_{
31,1} \eta_{3,12} \times\\
\times\mu_{12,31} \nu_{31}{-}2 \eta_{23,2} \eta_{2,23} \eta_{31,1} \eta_{3,12} \mu_{12,31} \nu_{31}{+}2 \eta_{23,1} \eta_{2,23} \eta_{31,2} \eta_{3,12} \mu_{12,31} \nu_{31}{+}\\
continued...
\end{array}
\end{equation*}
\begin{equation*}
\begin{array}{l}
...continued\\
+2 \eta_{23,2} \eta_{2,12} \eta_{31,1} \eta_{3,23} \mu_{12,31} \nu_{31}{+}2 \eta_{12,1} \eta_{3,23} \mu_{
23,12} \nu_{31}{+}2 \eta_{12,2} \eta_{2,31} \eta_{31,1} \eta_{3,23} \times\\
\times\mu_{23,12} \nu_{31}{-}2 \eta_{12,1} \eta_{2,31} \eta_{31,2} \eta_{3,23} \mu_{23,12} \nu_{31}{+}2 \eta_{12,1} \eta_{2,23} \eta_{31,2} \eta_{3,31} \mu_{23,12} \nu_{31}{+}\\
+2 \eta_{31,1} \eta_{3,31} \mu_{12,23} \mu_{23,12} \nu_{31}{-}2 \eta_{31,1} \eta_{3,23} \mu_{12,31} \mu_{23,12} \nu_{31}{+}2 \eta_{12,2} \eta_{2,23} \eta_{
31,1} \eta_{3,12}\times\\
\times \mu_{23,31} \nu_{31}{-}4 \eta_{12,1} \eta_{2,23} \eta_{31,2} \eta_{3,12} \mu_{23,31} \nu_{31}{+}2 \eta_{31,1} \eta_{3,23} \mu_{23,31} \nu_{31}{-}2 \eta_{12,2} \eta_{2,12} \times\\
\times\eta_{31,1} \eta_{3,23} \mu_{23,31} \nu_{31}{+}2 \eta_{12,1} \eta_{2,12} \eta_{31,2} \eta_{3,23} \mu_{23,31} \nu_{31}{-}2 \eta_{31,1} \eta_{3,12} \mu_{12,23} \mu_{23,31} \nu_{31}{-}\\
-4 \eta_{12,2} \eta_{23,1} \eta_{2,31} \eta_{3,23} \mu_{31,12} \nu_{31}{+}2 \eta_{12,1} \eta_{23,2} \eta_{2,31} \eta_{3,23} \mu_{31,12} \nu_{31}{+}2 \eta_{12,1} \eta_{3,31}\times\\
\times \mu_{31,12} \nu_{31}{+}2 \eta_{12,2} \eta_{23,1} \eta_{2,23} \eta_{3,31} \mu_{31,12} \nu_{31}{-}2 \eta_{12,1} \eta_{23,2} \eta_{2,23} \eta_{3,31} \mu_{31,12} \nu_{31}{-}\\
-2 \eta_{23,1} \eta_{3,31} \mu_{12,23} \mu_{31,12} \nu_{31}{+}2 \eta_{23,1} \eta_{3,23} \mu_{12,31} \mu_{31,12} \nu_{31}{-}2 \eta_{12,1} \eta_{3,23} \mu_{23,31}\times\\
\times \mu_{31,12} \nu_{31}{+}2 \eta_{12,2} \eta_{23,1} \eta_{2,31} \eta_{3,12} \mu_{31,23} \nu_{31}{+}2 \eta_{23,1} \eta_{3,31} \mu_{31,23} \nu_{31}{-}2 \eta_{12,2} \eta_{23,1} \times\\
\times\eta_{2,12} \eta_{3,31} \mu_{31,23} \nu_{31}{+}2 \eta_{12,1} \eta_{23,2} \eta_{2,12} \eta_{3,31} \mu_{31,23} \nu_{31}{-}2 \eta_{23,1} \eta_{3,12} \mu_{12,31} \mu_{31,23} \nu_{31}{-}\\
-2 \eta_{12,1} \eta_{3,31} \mu_{23,12} \mu_{31,23} \nu_{31}{+}2 \eta_{12,1} \eta_{3,12} \mu_{23,31} \mu_{31,23} \nu_{31}{-}2 \eta_{12,2} \eta_{3,12} \nu_{12} \nu_{31}{-}\\
-2 \eta_{23,2} \eta_{3,23} \nu_{12} \nu_{31}{-}2 \eta_{31,2} \eta_{3,31}\nu_{12} \nu_{31}{+}2 \eta_{23,2} \eta_{3,12} \mu_{12,23} \nu_{12} \nu_{31}{+}2 \eta_{31,2} \eta_{3,12} \mu_{12,31}\times\\
\times \nu_{12} \nu_{31}{+}2 \eta_{12,2} \eta_{3,23} \mu_{23,12} \nu_{12} \nu_{31}{+}2\eta_{31,2} \eta_{3,31} \mu_{12,23} \mu_{23,12} \nu_{12} \nu_{31}{-}2 \eta_{31,2} \eta_{3,23}\times\\
\times\mu_{12,31} \mu_{23,12} \nu_{12} \nu_{31}{+}2 \eta_{31,2} \eta_{3,23} \mu_{23,31} \nu_{12}\nu_{31}{-}2 \eta_{31,2} \eta_{3,12} \mu_{12,23} \mu_{23,31} \nu_{12} \nu_{31}{+}\\
+2 \eta_{12,2}\eta_{3,31} \mu_{31,12} \nu_{12} \nu_{31}{-}2 \eta_{23,2} \eta_{3,31} \mu_{12,23} \mu_{31,12}
\nu_{12} \nu_{31}{+}2 \eta_{23,2} \eta_{3,23} \mu_{12,31} \mu_{31,12}\times\\
\times \nu_{12} \nu_{31}{-}2 \eta_{12,2} \eta_{3,23} \mu_{23,31} \mu_{31,12} \nu_{12} \nu_{31}{+}2 \eta_{23,2} \eta_{3,31} \mu_{31,23} \nu_{12} \nu_{31}{-}2 \eta_{23,2} \eta_{3,12}\times\\
\times \mu_{12,31} \mu_{31,23} \nu_{12} \nu_{31}{-}2 \eta_{12,2} \eta_{3,31} \mu_{23,12} \mu_{31,23} \nu_{12} \nu_{31}{+}2 \eta_{12,2} \eta_{3,12} \mu_{23,31} \mu_{31,23} \nu_{12} \nu_{31}{-}\\
-\nu_{13} \nu_{31}{+}\eta_{12,2} \eta_{2,12}\nu_{13} \nu_{31}{+}\eta_{23,2} \eta_{2,23} \nu_{13} \nu_{31}{+}\eta_{2,31} \eta_{31,2} \nu_{13} \nu_{31}{-}2 \eta_{12,2} \eta_{2,23} \mu_{23,12}\times\\
\times \nu_{13} \nu_{31}{+}\mu_{12,23} \mu_{23,12}\nu_{13} \nu_{31}{-}\eta_{2,31} \eta_{31,2} \mu_{12,23} \mu_{23,12} \nu_{13} \nu_{31}\hspace{-.7mm}{-}2 \eta_{2,23} \eta_{31,2} \mu_{23,31} \nu_{13} \nu_{31}\hspace{-.7mm}{+}\\
+2 \eta_{2,12} \eta_{31,2} \mu_{12,23} \mu_{23,31} \nu_{13} \nu_{31}{-}2 \eta_{12,2} \eta_{2,31} \mu_{31,12} \nu_{13} \nu_{31}{+}2 \eta_{23,2} \eta_{2,31} \mu_{12,23} \mu_{31,12} \times\\
\times\nu_{13} \nu_{31}{+}\mu_{12,31} \mu_{31,12} \nu_{13} \nu_{31}{-}\eta_{23,2} \eta_{2,23} \mu_{12,31} \mu_{31,12} \nu_{13} \nu_{31}{+}2 \eta_{12,2}
\eta_{2,23} \mu_{23,31}\times\\
\times \mu_{31,12} \nu_{13} \nu_{31}{-}2 \mu_{12,23} \mu_{23,31} \mu_{31,12}\nu_{13} \nu_{31}{+}\mu_{23,31} \mu_{31,23} \nu_{13} \nu_{31}{-}\eta_{12,2} \eta_{2,12} \mu_{23,31} \times\\
\times\mu_{31,23} \nu_{13} \nu_{31}{-}2 \eta_{12,1} \eta_{2,12} \nu_{23} \nu_{31}{-}2 \eta_{23,1} \eta_{2,23} \nu_{23} \nu_{31}{-}2 \eta_{2,31} \eta_{31,1} \nu_{23} \nu_{31}{+}\\
+2 \eta_{23,1}\eta_{2,12} \mu_{12,23} \nu_{23} \nu_{31}{+}2 \eta_{2,12} \eta_{31,1} \mu_{12,31} \nu_{23}\nu_{31}{+}2 \eta_{12,1} \eta_{2,23} \mu_{23,12} \nu_{23} \nu_{31}{+}\\
+2 \eta_{2,31} \eta_{31,1}\mu_{12,23} \mu_{23,12} \nu_{23} \nu_{31}{-}2 \eta_{2,23} \eta_{31,1} \mu_{12,31} \mu_{23,12}\nu_{23} \nu_{31}{+}2 \eta_{2,23} \eta_{31,1} \mu_{23,31}\times\\
\times \nu_{23} \nu_{31}{-}2 \eta_{2,12} \eta_{31,1} \mu_{12,23} \mu_{23,31} \nu_{23} \nu_{31}{+}2 \eta_{12,1} \eta_{2,31} \mu_{31,12} \nu_{23} \nu_{31}{-}2 \eta_{23,1} \eta_{2,31} \times\\
\times\mu_{12,23} \mu_{31,12} \nu_{23} \nu_{31}{+}2 \eta_{23,1} \eta_{2,23} \mu_{12,31} \mu_{31,12} \nu_{23} \nu_{31}{-}2 \eta_{12,1} \eta_{2,23} \mu_{23,31} \mu_{31,12} \nu_{23} \nu_{31}{+}\\
+2 \eta_{23,1} \eta_{2,31} \mu_{31,23} \nu_{23} \nu_{31}{-}2 \eta_{23,1} \eta_{2,12} \mu_{12,31} \mu_{31,23} \nu_{23} \nu_{31}{-}2 \eta_{12,1} \eta_{2,31} \mu_{23,12} \mu_{31,23}\times\\
\times \nu_{23} \nu_{31}{+}2 \eta_{12,1} \eta_{2,12} \mu_{23,31} \mu_{31,23} \nu_{23} \nu_{31}{-}2 \nu_{12} \nu_{23} \nu_{31}{+}2 \mu_{12,23} \mu_{23,12} \nu_{12}\nu_{23} \nu_{31}{+}\\
+2 \mu_{12,31} \mu_{31,12} \nu_{12} \nu_{23} \nu_{31}{-}4 \mu_{12,23} \mu_{23,31} \mu_{31,12} \nu_{12} \nu_{23} \nu_{31}{+}2 \mu_{23,31} \mu_{31,23} \nu_{12} \nu_{23}\nu_{31}{-}\\
-\nu_{23} \nu_{32}{+}\mu_{12,23} \mu_{23,12} \nu_{23} \nu_{32}{+}\mu_{12,31} \mu_{31,12} \nu_{23} \nu_{32}{-}2 \mu_{12,23} \mu_{23,31} \mu_{31,12} \nu_{23} \nu_{32}{+}\\
+\mu_{23,31}\mu_{31,23} \nu_{23} \nu_{32}{+}[\eta_{1,31} \eta_{31,1}{+}\mu_{12,23} (2 \eta_{1,12} \eta_{31,1} \mu_{23,31}{-}2 \eta_{1,12} \eta_{23,1}{-}\\
-\eta_{1,31} \eta_{31,1} \mu_{23,12}{+}2 \eta_{1,31} \eta_{23,1} \mu_{31,12}){+}\eta_{1,23} (\eta_{23,1}{-}2 \eta_{31,1} \mu_{23,31}{-}\eta_{23,1} \mu_{12,31} \mu_{31,12}){+}\\
+\eta_{12,1} (\eta_{1,12}{-}2 \eta_{1,31} \mu_{31,12}{+}2 \eta_{1,23} \mu_{23,31} \mu_{31,12}{-}\eta_{1,12}\mu_{23,31} \mu_{31,23})] \nu_{23} \nu_{32}\}>0.
\end{array}
\end{equation*}

\subsection{Some explicit bounds}
By mathematical manipulations, the above necessary and sufficient conditions can give some particular indications on the technical constants. These indications we call {\it elastic bounds}. As a result of  mathematical treatments on some of the above conditions, they can be more than six but they just give a set of necessary, but {\it not} sufficient conditions for $V$ to be positive for any stress state. However, their interest is in the fact that such bounds are simple conditions that the elastic constants must necessarily satisfy.

\subsubsection{Bounds for the moduli}
Solving conditions (\ref{eq:triclinic1}) to (\ref{eq:triclinic6}) sequentially, we get immediately the well known bounds for the moduli:
\be
\label{eq:boundsmoduli}
E_i>0,\ \ G_{ij}>0\ \forall i,j=1,2,3.
\ee
The same result can be obtained if the order of the components of $\{\sigma\}$ is changed, i.e. if we organize differently the matrix $[S]$. As said above, the order of the components $\sigma_i$s in $\{\sigma\}$ is arbitrary. So, putting alternatively as first component of $\{\sigma\}$ all the $\sigma_i$s, we get again the result in eq. (\ref{eq:boundsmoduli})  from the condition on $M_1$. Other specific bounds can be obtained in the same way, this is done in the following Sections.

\subsubsection{Bounds for the Young's moduli and Poisson's ratios}
If we put at the first place successively $\sigma_1,\sigma_2$, as done above, then $\sigma_2,\sigma_3$ and $\sigma_3,\sigma_1$, the condition $M_2$ gives the general bound for the Poisson's coefficients
\be
\label{eq:poissons2}
1-\nu_{ij}\ \nu_{ji}>0\ \ \forall i,j=1,2,3.
\ee
Using again the reciprocity relations (\ref{eq:poissons1}) this result  can  be written also in the form
\be
 |\nu_{ij}|<\sqrt{\frac{E_i}{E_j}}\ \ \forall i,j=1,2,3.
\ee

The condition on $M_3$, eq. (\ref{eq:triclinic3}), gives a bound collecting together all the Poisson's rations:
\be
\label{eq:poissons3}
1-\nu_{12}\ \nu_{21}-\nu_{23}\ \nu_{32}-\nu_{31}\ \nu_{13}-2\nu_{12}\ \nu_{23}\ \nu_{31}>0.
\ee
Bounds (\ref{eq:poissons2}) and (\ref{eq:poissons3}), were already known, obtained using a mechanical approach, but only for orthotropic materials and with the constants expressed in the orthotropy frame, \cite{lempriere,jones}. The above relations prove that they are valid for all the materials. 

\subsubsection{Bounds for the shear moduli and the Chentsov's coefficients}
The procedure used in the previous Section can be repeated {\it verbatim} for finding some bounds for the shear moduli and the Chentsov's coefficients: if at the first two positions of $\{\sigma\}$ we put in turn $\sigma_4,\sigma_5$, then $\sigma_5,\sigma_6$ and finally $\sigma_6,\sigma_4$, using the reciprocity relations (\ref{eq:poissons1})$_2$, we get for the Chentsov's coefficients a general bound analogous to the one in eq. (\ref{eq:poissons2}), valid for the Poisson's ratios:
\be
\label{eq:chentscoef}
1-\mu_{ij,ki}\ \mu_{ki,ij}>0\ \ \forall i,j,k=1,2,3.
\ee
We can put these relations in a form similar to eq. (\ref{eq:poissons1}) thanks again to the reciprocity relations (\ref{eq:poissons1})$_2$:
\be
|\mu_{ij,ki}|<\sqrt{\frac{G_{ki}}{G_{ij}}}\ \forall i,j,k=1,2,3,
\ee
Moreover, ordering $\{\sigma\}$  with   $\sigma_4,\sigma_5,\sigma_6$ at the first three places and using once more eqs. (\ref{eq:poissons1})$_2$, from $M_3$ we obtain a bound for all the Chentsov's coefficients similar to the one concerning the Poisson's ratios, eq. (\ref{eq:poissons3}); the difference of sign  for the last term is due to the fact that in the definition of the Poisson's ratios there is a sign minus:
\be
\label{eq:chentsovcoefbis}
1-\mu_{12,23}\ \mu_{23,12}-\mu_{31,12}\ \mu_{12,31}-\mu_{23,31}\ \mu_{31,23}+2\mu_{23,31}\ \mu_{31,12}\ \mu_{12,23}>0.
\ee

\subsubsection{Bounds for the coefficients of mutual influence}
The same procedure can be used for obtaining some bounds concerning the coefficients of mutual influence of the first or of the second type; to this purpose, it is sufficient to put at the first two places of $\{\sigma\}$ respectively the nine couples of the type $\sigma_i,\sigma_p, i=1,2,3,\ p=4,5,6$. In such a way we will obtain the following bounds, concerning  the coefficients of mutual influence of the first type:
\be
\label{eq:mutinf1}
|\eta_{i,jk}|<\sqrt{\frac{G_{jk}}{E_i}}\ \forall i,j,k=1,2,3.
\ee
If we swap the positions in the nine couples, we will get the following bounds, concerning the coefficients of mutual influence of the second type:
\be
\label{eq:mutinf2}
|\eta_{ij,k}|<\sqrt{\frac{E_k}{G_{ij}}}\ \forall i,j,k=1,2,3.
\ee
Through the reciprocity relations (\ref{eq:poissons1})$_3$, from the above relations we get a general  bound for the coefficients of mutual influence  analogous to those already found for the Poisson's and Chentsov's coefficients:
\be
\label{eq:coefmutinf}
1-\eta_{i,jk}\ \eta_{jk,i}>0\ \forall i,j,k=1,2,3.
\ee
Using the reciprocity relations (\ref{eq:poissons1})$_{1,2}$  in the product of eqs. (\ref{eq:mutinf1}) and (\ref{eq:mutinf2}), we  obtain the following conditions linking together all the coefficients (Poisson's, Chen\-tsov's and of mutual influence of the first and of the second type):
\be
\label{eq:coeftous}
\eta^2_{i,jk}\ \eta^2_{ij,k}<\frac{\nu_{ki}}{\nu_{ik}}\frac{\mu_{ji,kj}}{\mu_{kj,ji}},\ \forall i,j,k=1,2,3.
\ee

Moreover, the condition on $M_4$, eq. (\ref{eq:triclinic4}), gives a bound for the Poisson's ratios and the coefficients of mutual influence of the first and second type:
\be
\label{eq:poissonmutinf1}
\begin{array}{l}
1-\nu_{12}\ \nu_{21}-\nu_{23}\ \nu_{32}-\nu_{31}\ \nu_{13}-\\
-2\nu_{12}\ \nu_{23}\ \nu_{31}+\\
+\nu_{12}\ \nu_{21}\ \eta_{3,23}\ \eta_{23,3}+\\
+\nu_{23}\ \nu_{32}\ \eta_{1,23}\ \eta_{23,1}+\\
+\nu_{31}\ \nu_{13}\ \eta_{2,23}\ \eta_{23,2}-\\
-\eta_{1,23}\left[\eta_{23,1}+2\nu_{12}(\eta_{23,2}+\eta_{23,3}\nu_{23})\right]-\\
-\eta_{2,23}\left[\eta_{23,2}+2\nu_{23}(\eta_{23,3}+\eta_{23,1}\nu_{31})\right]-\\
-\eta_{3,23}\left[\eta_{23,3}+2\nu_{31}(\eta_{23,1}+\eta_{23,2}\nu_{12})\right]>0.
\end{array}
\ee
Two  more bounds of the same type can be obtained if at the fourth place in $\{\sigma\}$ we put successively $\sigma_5$ and $\sigma_6$:
\be
\label{eq:poissonmutinf2}
\begin{array}{l}
1-\nu_{12}\ \nu_{21}-\nu_{23}\ \nu_{32}-\nu_{31}\ \nu_{13}-\\
-2\nu_{12}\ \nu_{23}\ \nu_{31}+\\
+\nu_{12}\ \nu_{21}\ \eta_{3,31}\ \eta_{31,3}+\\
+\nu_{23}\ \nu_{32}\ \eta_{1,31}\ \eta_{31,1}+\\
+\nu_{31}\ \nu_{13}\ \eta_{2,31}\ \eta_{31,2}-\\
-\eta_{1,31}\left[\eta_{31,1}+2\nu_{12}(\eta_{31,2}+\eta_{31,3}\nu_{23})\right]-\\
-\eta_{2,31}\left[\eta_{31,2}+2\nu_{23}(\eta_{31,3}+\eta_{31,1}\nu_{31})\right]-\\
-\eta_{3,31}\left[\eta_{31,3}+2\nu_{31}(\eta_{31,1}+\eta_{31,2}\nu_{12})\right]>0,
\end{array}
\ee
\be
\label{eq:poissonmutinf3}
\begin{array}{l}
1-\nu_{12}\ \nu_{21}-\nu_{23}\ \nu_{32}-\nu_{31}\ \nu_{13}-\\
-2\nu_{12}\ \nu_{23}\ \nu_{31}+\\
+\nu_{12}\ \nu_{21}\ \eta_{3,12}\ \eta_{12,3}+\\
+\nu_{23}\ \nu_{32}\ \eta_{1,12}\ \eta_{12,1}+\\
+\nu_{31}\ \nu_{13}\ \eta_{2,12}\ \eta_{12,2}-\\
-\eta_{1,12}\left[\eta_{12,1}+2\nu_{12}(\eta_{12,2}+\eta_{12,3}\nu_{23})\right]-\\
-\eta_{2,12}\left[\eta_{12,2}+2\nu_{23}(\eta_{12,3}+\eta_{12,1}\nu_{31})\right]-\\
-\eta_{3,12}\left[\eta_{12,3}+2\nu_{31}(\eta_{12,1}+\eta_{12,2}\nu_{12})\right]>0.
\end{array}
\ee

\subsection{Considerations on the triclinic case}
The results found above deserve some considerations:
\begin{itemize}
\item equations (\ref{eq:triclinic1}) to (\ref{eq:triclinic6}) constitute a set of independent necessary and sufficient conditions for $V$ to be positive for any stress state for triclinic materials, classes 1 and 2 of the Voigt's crystals classification, \cite{voigt,vannucci_libro};
\item they are valid also for any other anisotropic material when its technical constants are not referred to the existing material symmetry axes; for instance, they are valid also for an orthotropic material when its moduli are not given in the orthotropic frame;
\item the coefficients of mutual influence appear in $M_4$ to $M_6$ while the Chentsov's coefficients only in $M_5$ and $M_6$;
\item considering just the conditions for $M_1$ and $M_2$ and ordering in different ways the vector $\{\sigma\}$, not only the classical conditions stating the necessary positivity of the Young's and shear moduli, $E_i, G_{ij}$, are recovered, but also the  conditions (\ref{eq:poissons2}) on the Poisson's ratios, (\ref{eq:chentscoef}) on the Chentsov's coefficients and (\ref{eq:coefmutinf}) on the coefficients of mutual influence, are obtained. Among these ones, only eq. (\ref{eq:poissons2}) was already known, but erroneously considered valid uniquely for orthotropic materials;
\item a new set of  bounds, represented by eq. (\ref{eq:coeftous}) and linking together the Poisson's ratios, the Chentsov's coefficients and the coefficients of mutual influence, can be obtained combining the conditions on these last;
\item from the condition on $M_3$ it is possible to obtain not only the already known bound (\ref{eq:poissons3}), however also this one considered previously as valid only for orthotropic materials, but also, rearranging the order of $\{\sigma\}$, a similar, new bound concerning the Chentsov's coefficients, eq. (\ref{eq:chentsovcoefbis});
\item a set of three new bounds, eqs. (\ref{eq:poissonmutinf1}) to (\ref{eq:poissonmutinf3}), linking the Poisson's ratios with the coefficients of mutual influence can be obtained from $M_4$, also in this case reordering $\{\sigma\}$;
\item to the best knowledge of the author, bounds (\ref{eq:chentscoef}), (\ref{eq:coefmutinf}), (\ref{eq:coeftous}), (\ref{eq:chentsovcoefbis}), (\ref{eq:poissonmutinf1}), (\ref{eq:poissonmutinf2}) and (\ref{eq:poissonmutinf3}) are the first conditions concerning the Chentsov's coefficients and the coefficients of mutual influence;
\item all the bounds concerning the moduli, i.e. the $E_i$s and the $G_{ij}$s, measuring direct stiffnesses, stem from $M_1$, i.e. from a condition of the first order that can be applied to each one of the moduli simply by a suitable order of the $\sigma_i$s in $\{\sigma\}$; because  this strategy do not depend on the eventual elastic symmetries, this result is valid for any syngony and in any frame;
\item from conditions $M_2$ to $M_6$ we derive bounds concerning one more modulus for each $M_i$ and then the coefficients, i.e. dimensionless quantities expressing mechanical couplings: the Poisson's, the $\nu_{ij}$, and Chentsov's, the $\mu_{ij,ki}$,  effects and the coupling between shear stresses and normal strains, the $\eta_{i,jk}$, or between the normal stresses and shear strains, the $\eta_{ij,k}$.
\end{itemize}
In the next Section, the bounds for each elastic syngony are detailed. The presence of some material symmetry reduces and simplifies the expressions of the $M_i$s, especially of $M_5$ and $M_6$.

\section{Bounds for materials of  different elastic syngonies}
\subsection{Monoclinic materials}
Let us consider a monoclinic material with $x_3=0$ as plane of elastic symmetry (classes 3, 4 and 5 of the Voigt's crystal classification). In this case, cf. \cite{vannucci_libro}, 
 \be
 \label{eq:Smonoclinic}
 [S]=\left[
 \begin{array}{cccccc}
 S_{11}&S_{12}&S_{13}&0&0&S_{16}\\
 S_{12}&S_{22}&S_{23}&0&0&S_{26}\\
 S_{13}&S_{23}&S_{33}&0&0&S_{36}\\
 0&0&0&S_{44}&S_{45}&0\\
0&0&0&S_{45}&S_{55}&0\\
 S_{16}&S_{26}&S_{36}&0&0&S_{66}
 \end{array}
 \right]\rightarrow\ \begin{array}{l}\begin{array}{l}\eta_{i,23}=\eta_{23,i}=0,\\
 \eta_{i,31}=\eta_{31,i}=0,
 \end{array}\ \
   i=1,2,3,\\
 \begin{array}{l}
 \mu_{12,23}=\mu_{23,12}=0,\\
 \mu_{31,12}=\mu_{12,31}=0.
 \end{array}
 \end{array}
 \ee
Putting these conditions into eqs. (\ref{eq:triclinic1}) to  (\ref{eq:triclinic6}) gives:
\be
\begin{array}{c}
\begin{array}{c}
M_1>0 \rightarrow  \dfrac{1}{E_1}>0,\medskip\\
M_2>0 \rightarrow  \dfrac{1-\nu_{12}\ \nu_{21}}{E_1E_2}>0,\medskip\\
M_3>0 \rightarrow  \dfrac{1- \nu_{12}\ \nu_{21} -\nu_{23}\ \nu_{32}- \nu_{31} \ \nu_{13}    -2 \nu_{12}\ \nu_{23}\ \nu_{31}}{E_1E_2E_3}>0,\medskip\\
M_4>0 \rightarrow  \dfrac{1- \nu_{12}\ \nu_{21} -\nu_{23}\ \nu_{32}- \nu_{31} \ \nu_{13}    -2 \nu_{12}\ \nu_{23}\ \nu_{31}}{E_1E_2E_3G_{23}}>0,
\end{array}\medskip\\
\begin{array}{ll}
M_5>0 \rightarrow & \dfrac{1}{E_1E_2E_3G_{23}G_{31}}(1-\mu_{23,31}\ \mu_{31,23})\times\\
&\times(1- \nu_{12}\ \nu_{21} -\nu_{23}\ \nu_{32}- \nu_{31}\  \nu_{13}    -2 \nu_{12}\ \nu_{23}\ \nu_{31})>0,\medskip\\
M_6>0\rightarrow&\dfrac{1}{E_1E_2E_3G_{12}G_{23}G_{31}}(1-\mu_{23,31}\ \mu_{31,23})\times\\
&\times\left\{1- \nu_{12}\ \nu_{21} -\nu_{23}\ \nu_{32}- \nu_{31} \ \nu_{13}    -2 \nu_{12}\ \nu_{23}\ \nu_{31}+\right.\\
&+\nu_{12}\ \nu_{21}\ \eta_{12,3}\ \eta_{3,12}+\nu_{23}\ \nu_{32}\ \eta_{12,1}\ \eta_{1,12}+ \nu_{31} \ \nu_{13}\ \eta_{12,2}\ \eta_{2,12}-\\
&-\eta_{1,12}\left[\eta_{12,1}+2\nu_{12}(\eta_{12,2}+\eta_{12,3}\ \nu_{23})\right]-\\
&-\eta_{2,12}\left[\eta_{12,2}+2\nu_{23}(\eta_{12,3}+\eta_{12,1}\ \nu_{31})\right]-\\
&\left.-\eta_{3,12}\left[\eta_{12,3}+2\nu_{31}(\eta_{12,1}+\eta_{12,2}\ \nu_{12})\right]\right\}>0.
\end{array}
\end{array}
\ee

The simplification in the expressions of $M_5$ and $M_6$ given by the existence of a plane of elastic symmetry is remarkable. To notice that the coefficients of mutual influence now appear only in $M_6$. From these six necessary and sufficient conditions for the positivity of $V$ we get easily, proceeding from $M_1$ to $M_6$, the following {\it complete set of independent bounds}, in the sense that they are equivalent to the previous conditions $M_1$ to $M_6$, separately for the moduli and the coefficients:
\be
\begin{array}{l}
E_i>0, \ \ G_{ij}>0\ \forall i,j=1,2,3,\smallskip\\
1-\nu_{12}\ \nu_{21}>0,\smallskip\\
1- \nu_{12}\ \nu_{21} -\nu_{23}\ \nu_{32}- \nu_{31} \ \nu_{13}    -2 \nu_{12}\ \nu_{23}\ \nu_{31}>0,\smallskip\\
1-\mu_{23,31}\ \mu_{31,23}>0,\smallskip\\
1- \nu_{12}\ \nu_{21} -\nu_{23}\ \nu_{32}- \nu_{31} \ \nu_{13}    -2 \nu_{12}\ \nu_{23}\ \nu_{31}+\\
+\nu_{12}\ \nu_{21}\ \eta_{12,3}\ \eta_{3,12}+\nu_{23}\ \nu_{32}\ \eta_{12,1}\ \eta_{1,12}+ \nu_{31} \ \nu_{13}\ \eta_{12,2}\ \eta_{2,12}-\\
-\eta_{1,12}\left[\eta_{12,1}+2\nu_{12}(\eta_{12,2}+\eta_{12,3}\ \nu_{23})\right]-\\
-\eta_{2,12}\left[\eta_{12,2}+2\nu_{23}(\eta_{12,3}+\eta_{12,1}\ \nu_{31})\right]-\\
-\eta_{3,12}\left[\eta_{12,3}+2\nu_{31}(\eta_{12,1}+\eta_{12,2}\ \nu_{12})\right]>0.
\end{array}
\ee
Also in this case changing the order of the $\sigma_i$s in $\{\sigma\}$ will give some other specific bounds for the coefficients, namely, changing the order of $\sigma_1,\sigma_2,\sigma_3$ will give again eq. (\ref{eq:poissons2}), while putting $\sigma_4,\sigma_5,\sigma_6$ at the first three places  gives the additional bounds
\be
\label{eq:morebounds}
1-\eta_{1,12}\ \eta_{12,1}>0,\ \ 1-\eta_{2,12}\ \eta_{12,2}-\nu_{12}\ \nu_{21}-\eta_{1,12}(\eta_{12,1}+2\eta_{12,2}\ \nu_{12})>0.
\ee

\subsection{Orthotropic materials}
For an orthotropic material, like crystals of the {\it orthorombic syngony}, classes 6,7 and 8 of  the Voigt's crystals classification,  with $x_1=0, x_2=0,x_3=0$ as planes of elastic symmetry, the matrix $[S]$ is
 \be
 \label{eq:Sorthotropic}
 [S]=\left[
 \begin{array}{cccccc}
 S_{11}&S_{12}&S_{13}&0&0&0\\
 S_{12}&S_{22}&S_{23}&0&0&0\\
 S_{13}&S_{23}&S_{33}&0&0&0\\
 0&0&0&S_{44}&0&0\\
0&0&0&0&S_{55}&0\\
 0&0&0&0&0&S_{66}
 \end{array}
 \right]\rightarrow\ 
 \begin{array}{l}
 \eta_{i,jk}=\eta_{jk,i}=0,\\
 \mu_{ij,jk}=\mu_{jk,ij}=0,
 \end{array}
 \ i,j,k=1,2,3.
 \ee
 Injecting these conditions into eqs. (\ref{eq:triclinic1}) to  (\ref{eq:triclinic6}) gives, 
\be
\begin{array}{c}
M_1>0 \rightarrow  \dfrac{1}{E_1}>0,\medskip\\
M_2>0 \rightarrow  \dfrac{E_1-\nu_{12}^2E_2}{E_1E_2}>0,\medskip\\
M_3>0 \rightarrow  \dfrac{1- \nu_{12}\ \nu_{21} -\nu_{23}\ \nu_{32}- \nu_{31} \ \nu_{13}    -2 \nu_{12}\ \nu_{23}\ \nu_{31}}{E_1E_2E_3}>0,\medskip\\
M_4>0 \rightarrow  \dfrac{1- \nu_{12}\ \nu_{21} -\nu_{23}\ \nu_{32}- \nu_{31} \ \nu_{13}    -2 \nu_{12}\ \nu_{23}\ \nu_{31}}{E_1E_2E_3G_{23}}>0,\medskip\\
M_5>0 \rightarrow  \dfrac{
1- \nu_{12}\ \nu_{21} -\nu_{23}\ \nu_{32}- \nu_{31} \ \nu_{13}  -2 \nu_{12}\ \nu_{23}\ \nu_{31}}{E_1E_2E_3G_{23}G_{31}}>0,\medskip\\
M_6>0\rightarrow\dfrac{1- \nu_{12}\ \nu_{21} -\nu_{23}\ \nu_{32}- \nu_{31} \ \nu_{13}  -2 \nu_{12}\ \nu_{23}\ \nu_{31}}{E_1E_2E_3G_{12}G_{23}G_{31}}>0.
\end{array}
\ee

Proceeding as before we get the complete set of independent bounds
\be
\begin{array}{l}
E_i>0, \ \ G_{ij}>0\ \forall i,j=1,2,3,\\
1-\nu_{12}\ \nu_{21}>0,\\
1- \nu_{12}\ \nu_{21} -\nu_{23}\ \nu_{32}- \nu_{31} \ \nu_{13}    -2 \nu_{12}\ \nu_{23}\ \nu_{31}>0,
\end{array}
\ee
while a change in the order of $\sigma_1,\sigma_2,\sigma_3$ still gives eqs. (\ref{eq:poissons2}). Of course, the results for the orthotropic case can also be obtained directly from those found in the previous Section for monoclinic materials, simply putting $\mu_{23,31}=\eta_{1,12}=\eta_{2,12}=\eta_{3,12}=0$. The last bounds were already known, but it was not clear, in the literature, whether or not they constitue a complete set of independent bounds for the elastic constants; normally, cf. e.g. \cite{jones}, the bounds that are written for the orthotropy case are the above ones and eqs. (\ref{eq:poissons2}). The result found here shows that this should be redundant for the positivity of $V$.

\subsection{Materials of the trigonal syngony}

Let us consider now materials having a 3-fold rotational symmetry, i.e. the covering operation, bringing two equivalent directions to coincidence, is a rotation of $2/3\pi$ about, say, the axis $x_3$; in this case
\be
 \label{eq:Strigonal}
 \begin{array}{ll}
 [S]=&\left[
 \begin{array}{cccccc}
 S_{11}&S_{12}&S_{13}&S_{14}&S_{15}&0\\
 S_{12}&S_{11}&S_{13}&-S_{14}&-S_{15}&0\\
 S_{13}&S_{13}&S_{33}&0&0&0\\
 S_{14}&-S_{14}&0&S_{44}&0&-\sqrt{2}S_{15}\\
S_{15}&-S_{15}&0&0&S_{44}&\sqrt{2}S_{14}\\
 0&0&0&-\sqrt{2}S_{15}&\sqrt{2}S_{14}&S_{11}-S_{12}
 \end{array}
 \right]\rightarrow\medskip\\ 
& \begin{array}{l}
E_2=E_1,\ \ G_{31}=G_{23},\ G_{12}=\dfrac{E_1}{2(1+\nu_{12})},\\
\nu_{12}=\nu_{21},\ \nu_{32}=\nu_{31},\ \nu_{23}=\nu_{13},\\
 \eta_{i,12}=\eta_{12,i}=0,\  i=1,2,3,\\
 \eta_{2,23}=-\eta_{1,23},\ \eta_{23,2}=-\eta_{23,1},\\
  \eta_{2,31}=-\eta_{1,31},\ \eta_{31,2}=-\eta_{31,1},\\
\eta_{3,23}=\eta_{23,3}=\eta_{3,31}=\eta_{31,3}=0,\\
 \mu_{23,31}=\mu_{31,23}=0,\\
 \mu_{12,23}=-2\eta_{1,31},\ \mu_{23,12}=-\dfrac{\eta_{31,1}}{1+\nu_{12}},\\
 \mu_{31,12}=\dfrac{\eta_{23,1}}{1+\nu_{12}},\ \mu_{12,31}=2\eta_{1,23}.
 \end{array}
 \end{array}
 \ee
These materials belong to classes 12 and 13 of the Voigt's classification and their elastic behavior depends upon seven independent elastic parameters. Writing the conditions $M_1$ to $M_6$ for such materials gives

\be
\begin{array}{c}
M_1>0 \rightarrow  \dfrac{1}{E_1}>0,\medskip\\
M_2>0 \rightarrow  \dfrac{1-\nu_{12}^2}{E_1^2}>0,\medskip\\
M_3>0 \rightarrow  \dfrac{(1+ \nu_{12})(1-\nu_{12} -2 \nu_{13}\ \nu_{31})}{E_1^2E_3}>0,\medskip\\
M_4>0 \rightarrow  \dfrac{(1+ \nu_{12}-2\eta_{1,23}\ \eta_{23,1})(1-\nu_{12} -2 \nu_{13}\ \nu_{31})}{E_1^2E_3G_{23}}>0,\medskip\\
M_5>0 \rightarrow  \dfrac{(1+ \nu_{12} -2\eta_{1,23}\ \eta_{23,1}-2\eta_{1,31}\ \eta_{31,1})(1-\nu_{12} -2 \nu_{13}\ \nu_{31})}{E_1^2E_3G_{23}^2}>0,\medskip\\
M_6>0\rightarrow   \dfrac{(1+ \nu_{12} -2\eta_{1,23}\ \eta_{23,1}-2\eta_{1,31}\ \eta_{31,1})^2(1-\nu_{12} -2 \nu_{13}\ \nu_{31})}{E_1^3E_3G_{23}^2}>0.
\end{array}
\ee
It is apparent that the condition $M_6$ is redundant. From the above relations we obtain the following complete set of seven bounds:
\be
\label{eq:boundtrigonal}
\begin{array}{l}
E_1>0,\ E_3>0,\ G_{23}>0,\\
1-\nu_{12}^2>0,\\
1-\nu_{12} -2 \nu_{13}\ \nu_{31}>0,\\
1+ \nu_{12}-2\eta_{1,23}\eta_{23,1}>0,\\
1+ \nu_{12} -2\eta_{1,23}\ \eta_{23,1}-2\eta_{1,31}\ \eta_{31,1}>0.
\end{array}
\ee

In the trigonal syngony, we can find also materials with only six independent parameters. This happens when to the rotational symmetry we add a mirror symmetry; such materials belong to classes 9, 10 and 11 of the Voigt's classification. When plane $x_1=0$ is the mirror, then also $S_{15}{=}0\rightarrow \eta_{1,31}=\eta_{31,1}=0$ and by eq. (\ref{eq:Strigonal}) it is also $S_{25}=0\rightarrow\eta_{2,31}=\eta_{31,2}=0$ and $S_{46}=0\rightarrow\mu_{12,23}=\mu_{23,12}=0$. The only changes in eq. (\ref{eq:Strigonal}) concern $M_5$ and $M_6$, where the term $-2\eta_{1,31}\ \eta_{31,1}$ disappears and consequently the last of bounds (\ref{eq:boundtrigonal})  becomes redundant.
If the mirror is the plane $x_2=0$, then $S_{14}{=}0\rightarrow \eta_{1,23}=\eta_{23,1}=0$ and again by eq. (\ref{eq:Strigonal}) it is also $S_{24}=0\rightarrow\eta_{2,23}=\eta_{23,2}=0$ and $S_{56}=0\rightarrow\mu_{12,31}=\mu_{31,12}=0$. In this case, the terms $-2\eta_{1,23}\ \eta_{23,1}$ disappear from $M_4,M_5$ and $M_6$ in eq. (\ref{eq:Strigonal}) while bounds (\ref{eq:boundtrigonal}) reduce to
\be
\begin{array}{l}
E_1>0,\ E_3>0,\ G_{23}>0,\\
1-\nu_{12}^2>0,\\
1-\nu_{12} -2 \nu_{13}\ \nu_{31}>0,\\
1+ \nu_{12} -2\eta_{1,31}\ \eta_{31,1}>0.
\end{array}
\ee

\subsection{Materials of the tetragonal syngony}
Belong to the tetragonal syngony materials having a 4-fold rotational symmetry; in this case, the covering operation is a rotation of $\pi/2$. Supposing again that the axis $x_3$ is the symmetry axis, then
\be
 \label{eq:Stetragonal}
 \begin{array}{ll}
 [S]=&\left[
 \begin{array}{cccccc}
 S_{11}&S_{12}&S_{13}&0&0&S_{16}\\
 S_{12}&S_{11}&S_{13}&0&0&-S_{16}\\
 S_{13}&S_{13}&S_{33}&0&0&0\\
 0&0&0&S_{44}&0&0\\
 0&0&0&0&S_{44}&0\\
 S_{16}&-S_{16}&0&0&0&S_{66}
 \end{array}
 \right]\rightarrow\medskip\\ 
& \begin{array}{l}
E_2=E_1,\ \ G_{31}=G_{23},\\
\nu_{12}=\nu_{21},\ \nu_{32}=\nu_{31},\ \nu_{23}=\nu_{13},\\
 \eta_{i,23}= \eta_{23,i}=\eta_{i,31}= \eta_{31,i}=0,\ i=1,2,3,\\
\eta_{3,23}=\eta_{23,3}=\eta_{3,31}=\eta_{31,3}=0,\\
\eta_{2,12}=-\eta_{1,12},\ \eta_{12,2}=-\eta_{12,1},\\
\eta_{3,12}=\eta_{12,3}=0,\\
 \mu_{ij,ki}=\mu_{ki,ij}=0, \ i,j,k=1,2,3.
 \end{array}
 \end{array}
 \ee
These materials belong to classes 17, 18 and 20 of the Voigt's classification and their elastic behavior also depends upon seven independent elastic parameters. The conditions $M_1$ to $M_6$ for such materials are
\be
\label{eq:Mitetragonal}
\begin{array}{c}
M_1>0 \rightarrow  \dfrac{1}{E_1}>0,\medskip\\
M_2>0 \rightarrow  \dfrac{1-\nu_{12}^2}{E_1^2}>0,\medskip\\
M_3>0 \rightarrow  \dfrac{(1+ \nu_{12})(1-\nu_{12} -2 \nu_{13}\ \nu_{31})}{E_1^2E_3}>0,\medskip\\
M_4>0 \rightarrow  \dfrac{(1+ \nu_{12})(1-\nu_{12} -2 \nu_{13}\ \nu_{31})}{E_1^2E_3G_{23}}>0,\medskip\\
M_5>0 \rightarrow  \dfrac{(1+ \nu_{12})(1-\nu_{12} -2 \nu_{13}\ \nu_{31})}{E_1^2E_3G_{23}^2}>0,\medskip\\
M_6>0\rightarrow   \dfrac{(1-\nu_{12} -2 \nu_{13}\ \nu_{31})(1+\nu_{12}-2\eta_{1,12}\ \eta_{12,1})}{E_1^2E_3G_{12}G_{23}^2}>0.
\end{array}
\ee
In this case, $M_5$ is redundant. The above relations reduce to the following complete set of seven bounds:
\be
\label{eq:boundtetragonal}
\begin{array}{l}
E_1>0,\ E_3>0,\ G_{12}>0,\ G_{23}>0,\\
1-\nu_{12}^2>0,\\
1-\nu_{12} -2 \nu_{13}\ \nu_{31}>0,\\
1+ \nu_{12}-2\eta_{1,12}\ \eta_{12,1}>0.
\end{array}
\ee

Also for the tetragonal syngony there are materials depending upon six parameters. %This happens when the plane $x_3=0$ is a plane of material symmetry. 
These materials belong to  classes 14, 15, 16 and 19 of the Voigt's classification and actually they  coincide with   orthotropic materials having the same properties along the directions $x_1$ and $x_2$. For these materials,  $S_{16}=0\rightarrow\eta_{1,12}=\eta_{12,1}=0$. In this circumstance, the only change in eq. (\ref{eq:Mitetragonal}) concerns $M_6$, where the term $-2\eta_{1,12}\ \eta_{12,1}$ disappears. Then, $M_6$ becomes redundant and bounds (\ref{eq:boundtetragonal}) reduce to only six:
\be
\begin{array}{l}
E_1>0,\ E_3>0,\ G_{12}>0,\ G_{23}>0,\\
1-\nu_{12}^2>0,\\
1-\nu_{12} -2 \nu_{13}\ \nu_{31}>0.
\end{array}
\ee

\subsection{Materials of the hexagonal syngony}
The case of the hexagonal syngony is that of materials having a 6-fold rotational symmetry: the covering operation corresponds to a rotation of $\pi/3$ about, say, axis $x_3$. In this case
\be
 \label{eq:Shexagonal}
 \begin{array}{ll}
 [S]=&\left[
 \begin{array}{cccccc}
 S_{11}&S_{12}&S_{13}&0&0&0\\
 S_{12}&S_{11}&S_{13}&0&0&0\\
 S_{13}&S_{13}&S_{33}&0&0&0\\
 0&0&0&S_{44}&0&0\\
0&0&0&0&S_{44}&0\\
 0&0&0&0&0&S_{11}-S_{12}
 \end{array}
 \right]\rightarrow\medskip\\
& \begin{array}{l}
E_2=E_1,\ \ G_{31}=G_{23},\ G_{12}=\dfrac{E_1}{2(1+\nu_{12})},\\
\nu_{12}=\nu_{21},\ \nu_{32}=\nu_{31},\ \nu_{23}=\nu_{13},\\
 \eta_{i,jk}=\eta_{jk,i}=0,\ \mu_{ij,jk}=\mu_{jk,ij}=0, \ i,j,k=1,2,3.
 \end{array}
 \end{array}
 \ee
Crystals  of classes 21 to 27 of the Voigt's classification are of this type, but not only. The above elastic matrix is also typical of {\it transversely isotropic materials}, that elastically are equivalent to the hexagonal syngony and  do not exist as crystals. For all these materials, the conditions $M_1$ to $M_6$ reads like
\be
\label{eq:Mihexagonal}
\begin{array}{c}
M_1>0 \rightarrow  \dfrac{1}{E_1}>0,\medskip\\
M_2>0 \rightarrow  \dfrac{1-\nu_{12}^2}{E_1^2}>0,\medskip\\
M_3>0 \rightarrow  \dfrac{(1+ \nu_{12})(1-\nu_{12} -2 \nu_{13}\ \nu_{31})}{E_1^2E_3}>0,\medskip\\
M_4>0 \rightarrow  \dfrac{(1+ \nu_{12})(1-\nu_{12} -2 \nu_{13}\ \nu_{31})}{E_1^2E_3G_{23}}>0,\medskip\\
M_5>0 \rightarrow  \dfrac{(1+ \nu_{12})(1-\nu_{12} -2 \nu_{13}\ \nu_{31})}{E_1^2E_3G_{23}^2}>0,\medskip\\
M_6>0\rightarrow   \dfrac{(1+ \nu_{12})^2(1-\nu_{12} -2 \nu_{13}\ \nu_{31})}{E_1^3E_3G_{23}^2}>0.
\end{array}
\ee
It is evident that conditions $M_5$ and $M_6$ are now redundant. From the above relations, we get the complete set of five bounds
\be
\begin{array}{l}
E_1>0,\ E_3>0,\ G_{23}>0,\\
1-\nu_{12}^2>0,\\
1-\nu_{12} -2 \nu_{13}\ \nu_{31}>0.
\end{array}
\ee

\subsection{Materials of the cubic syngony}
The materials of the cubic syngony are actually orthotropic materials with the supplementary property of having the same properties along the symmetry axes. They belong to classes 28 to 32 of the Voigt's classification and have elastic matrix 
 \be
 \label{eq:Scubic}
 [S]=\left[
 \begin{array}{cccccc}
 S_{11}&S_{12}&S_{12}&0&0&0\\
 S_{12}&S_{11}&S_{12}&0&0&0\\
 S_{11}&S_{12}&S_{11}&0&0&0\\
 0&0&0&S_{44}&0&0\\
0&0&0&0&S_{44}&0\\
 0&0&0&0&0&S_{44}
 \end{array}
 \right]\rightarrow\ 
 \begin{array}{l}
 \begin{array}{l}
 E_1=E_2=E_3,\ G_{12}=G_{23}=G_{31},\\
 \nu_{12}=\nu_{21}=\nu_{23}=\nu_{32}=\nu_{31}=\nu_{13},
 \end{array}\\
 \begin{array}{l}
 \eta_{i,jk}=\eta_{jk,i}=0,\\
 \mu_{ij,jk}=\mu_{jk,ij}=0,
 \end{array}
 \ i,j,k=1,2,3.
 \end{array}
 \ee
The conditions $M_1$ to $M_6$ are
\be
\label{eq:Micubic}
\begin{array}{c}
M_1>0 \rightarrow  \dfrac{1}{E_1}>0,\medskip\\
M_2>0 \rightarrow  \dfrac{1-\nu_{12}^2}{E_1^2}>0,\medskip\\
M_3>0 \rightarrow  \dfrac{(1+ \nu_{12})^2(1-2\nu_{12})}{E_1^3}>0,\medskip\\
M_4>0 \rightarrow  \dfrac{(1+ \nu_{12})^2(1-2\nu_{12})}{E_1^3G_{12}}>0,\medskip\\
M_5>0 \rightarrow  \dfrac{(1+ \nu_{12})^2(1-2\nu_{12})}{E_1^3G_{12}^2}>0,\medskip\\
M_6>0\rightarrow   \dfrac{(1+ \nu_{12})^2(1-2\nu_{12})}{E_1^3G_{12}^3}>0.
\end{array}
\ee
Conditions $M_5$ and $M_6$ are evidently redundant. Finally, we obtain a complete set of four bounds:
\be
E_1>0,\ G_{12}>0,\ -1<\nu_{12}<\frac{1}{2}.
\ee

\subsection{Isotropic materials}
Evidently, the last result when applied to isotropic materials, having the supplementary condition 
\be
G_{12}=\frac{E_1}{2(1+\nu_{12})},
\ee
gives the well known bounds  (we put as usual $E_i=E,\nu_{ij}=\nu\ \forall i,j=1,2,3$)
\be
E>0,\ -1<\nu<\frac{1}{2}.
\ee

 \subsection{Remarks on the bounds for  materials with elastic symmetries}
On the base of the results found in the previous Sections, some general considerations can be drawn:
\begin{itemize}
\item the presence of material symmetries drastically simplifies the expressions of conditions $M_5$ and $M_6$;
\item in some cases, one or more of these conditions are redundant;
\item from conditions $M_1$ to $M_6$, that, e recall, are necessary and sufficient conditions for the positivity of $V$, we van obtain a complete set of bounds for each syngony, i.e. a set of conditions on the technical constants that are equivalent to conditions $M_1$ to $M_6$, but having a more direct and explicit expression in terms of the technical constants;
\item in particular, we obtain separately bounds (of positivity) for the moduli $E_i$s and $G_{ij}$s, $i,j=1,2,3$ and for the coefficients;
\item the number of these bounds varies with the syngony: it is of ten for monoclinic materials,  eight for the case of orthotropy, seven for the trigonal and tetragonal syngony, six for the same syngonies with six independent constants, five for the hexagonal syngony and for transversely isotropic materials, four for the cubic syngony and  finally three for isotropy;
\item it is however possible to write supplementary bounds in each case, which express specific conditions for some of the elastic parameters, like those in eq. (\ref{eq:morebounds});
\item for the orthorombic, tetragonal with six constants, hexagonal, cubic syngonies, and of course for transverse isotropy and isotropy, there are no bounds concerning the coefficients of mutual influence and the Chentsov's coefficients (unless their elastic matrix is known in a frame whose axes do not coincide with the symmetry axes; in that case, the results are the same of the general, triclinic case).
\end{itemize}

\section{Conclusion}
The use of a rather forgotten theorem of linear algebra has allowed to obtain general and complete results for the bounds of the technical constants in linear elasticity. A first result is the fact that the necessary and sufficient conditions for the elastic energy $V$ to be positive for any possible state of strain or stress is of six, at most. The use of the reciprocity relations regarding the Poisson's ratio, the Chentsov's coefficients and the coefficients of mutual influence makes it possible to obtain conditions for the positivity of $V$  that have all the same mathematical structure. They are expressed by the product of two functions: one depending only upon the moduli, the other just on the coefficients. Though in the general case of {\it complete anisotropy} (i.e. triclinic materials or materials with some elastic symmetry but with the elastic matrix given in a frame different from the  one of the material symmetry) the conditions have a very long expression, it is nevertheless possible to obtain specific, simple bounds operating on such conditions. This has allowed to show, for instance, that some previously bounds considered as valid for orthotropic materials are in reality valid for all the materials, also for the completely anisotropic ones. When the procedure is applied to materials having some specific elastic symmetry, the expression of the conditions for the positivity of $V$ is greatly simplified and some of the conditions, in some cases, become redundant. In these situations, it is possible and easy to transform the six necessary and sufficient conditions for the positivity of $V$ in a set of equivalent bounds, expressing more directly some conditions on the moduli or on the coefficients, but separately: some bounds concern exclusively the moduli, some other only the coefficients. The number of these bounds varies with the syngony, from 8 for the monoclinic case, to 2 for isotropy and they constitute a complete set of bounds, in the sense that they constitute a set of necessary and sufficient conditions for the positivity of $V$. Finally, supplementary bounds, namely concerning the coefficients, can be written.

\bibliography{Biblio}   % name your BibTeX data base

\end{document}